\documentclass[lettersize,journal]{IEEEtran}
\usepackage{amsmath,amsfonts}
\usepackage{algorithmic}
\usepackage{algorithm}
\usepackage{array}
\usepackage[caption=false,font=footnotesize,labelfont=rm,textfont=rm]{subfig}
\usepackage{textcomp}
\usepackage{stfloats}
\usepackage{url}
\usepackage{verbatim}
\usepackage{graphicx}
\usepackage{cite}
\usepackage{multirow}
\usepackage {xcolor}
\usepackage[normalem]{ulem}
\useunder{\uline}{\ul}{}
\usepackage{makecell}
\usepackage{hyperref}

\usepackage{etoolbox}
\makeatletter
\patchcmd{\@makecaption}
  {\scshape}
  {}
  {}
  {}
\makeatother

\newcommand{\ie}{\textit{i.e.}, }
\newcommand{\etal}{\textit{et al. }}

\newcommand{\la}[1]{{\color[rgb]{0,0,0}#1}}

\newcommand{\laa}[1]{{\color[rgb]{0,0,0}#1}}
\newcommand{\lab}[1]{{\color[rgb]{0,0,0}#1}}

\newcommand{\an}[1]{{\color{black}#1}}
\newcommand{\ana}[1]{{\color{black}#1}}
\newcommand{\anb}[1]{{\color{black}#1}}

\newcommand{\anf}[1]{{\color{black}#1}}

\begin{document}














\title{Multimodal Action Quality Assessment}

\author{Ling-An Zeng and Wei-Shi Zheng
\IEEEcompsocitemizethanks{\IEEEcompsocthanksitem
Ling-An Zeng is with the School of Artificial Intelligence, Sun Yat-sen University, Zhuhai, Guangdong 519082, China (e-mail: zenglan3@mail2.sysu.edu.cn).
\IEEEcompsocthanksitem
Wei-Shi Zheng is with the School of Computer Science and Engineering, Sun Yat-sen University, Guangzhou, Guangdong 510275, China, and with Guangdong Key Laboratory of Information Security Technology, Sun Yat-sen University, Guangzhou, Guangdong 510275, China, and also with the Key Laboratory of Machine Intelligence and Advanced Computing, Sun Yat-sen University, Ministry of Education, Guangzhou, Guangdong 510275, China (e-mail: wszheng@ieee.org /zhwshi@mail.sysu.edu.cn).
\IEEEcompsocthanksitem Wei-Shi Zheng is the corresponding author.
\IEEEcompsocthanksitem}}

\maketitle

\begin{abstract}
Action quality assessment (AQA) is to assess how well an action is performed. Previous works perform modelling by only the use of visual information, ignoring audio information. We argue that although AQA is highly dependent on visual information, the audio is useful complementary information for improving the score regression accuracy, especially for sports with background music, such as figure skating and rhythmic gymnastics.  \laa{To leverage multimodal information for AQA, \ie RGB, optical flow and audio information, we propose a Progressive Adaptive Multimodal Fusion Network (PAMFN) that separately models modality-specific information and mixed-modality information.} \laa{Our model consists of} with three modality-specific branches that independently explore  modality-specific information and a mixed-modality branch that progressively aggregates the modality-specific information from the modality-specific branches. \la{To build the bridge between modality-specific branches and the mixed-modality branch, three novel modules are proposed. First, \laa{a Modality-specific Feature Decoder module} is designed to selectively transfer modality-specific information to the mixed-modality branch. Second, when exploring the interaction between modality-specific information, we argue that using an invariant multimodal fusion policy may lead to suboptimal results, \laa{so as to take the potential diversity in different parts of an action into consideration}. Therefore, an Adaptive Fusion Module is proposed to learn adaptive multimodal fusion policies in different parts of an action. This module consists of several FusionNets for exploring different multimodal fusion strategies and a PolicyNet for deciding which FusionNets are enabled. Third, a module called Cross-modal Feature Decoder is designed to transfer cross-modal features generated by Adaptive Fusion Module to the mixed-modality branch.} Our extensive experiments validate the efficacy of \lab{the proposed method, and our method} achieves state-of-the-art performance on two public datasets. \anf{Code is available at \href{https://github.com/qinghuannn/PAMFN}{https://github.com/qinghuannn/PAMFN.}
}
\end{abstract}


\section{Introduction}
\label{sec:intro}
Action quality assessment (AQA) is the task of assessing how well an action is performed and is usually modeled as a score regression problem. Different from tasks such as action recognition and action localization, AQA is a fine-grained action understanding task, which not only needs to recognize actions but also to understand the subtle differences among actions. For instance, an insufficient leg lift angle in a leg-lifting action does not significantly affect action recognition but results in substandard actions and penalties~\cite{zeng2020hybrid}. AQA has many important real-world applications, including in sports~\cite{pirsiavash2014assessing, xu2022likert, parmar2017learning, xu2019learning, tang2020uncertainty, pan2019action, gao2020asymmetric, zeng2020hybrid, yu2021group, liu2021towards, nagai2021action,nekoui2021eagle, pan2021adaptive, zhang2022semi, bai2022action, Xu_2022_CVPR},  surgical training \cite{zia2018video, pan2019action, gao2020asymmetric, pan2021adaptive} and other fields \cite{doughty2018s, doughty2019pros, li2019manipulation, parmar2022domain}.

Audio usually serves as complementary information in action recognition \cite{owens2018audio, xiao2020audiovisual, gao2020listen, Shi_2021_ICCV, Alfasly_2022_CVPR} and action localization\cite{liu2019completeness, wu2019dual, lee2020cross, Xia_2022_CVPR, Jiang_2022_CVPR}, and has been shown to significantly improve performance. Similarly, although AQA is highly dependent on visual information, audio can be used as complementary information to improve the score regression accuracy, especially in sports with background music. \ana{In skating and rhythmic gymnastics, athletes need to perform actions to the rhythm of music, and a mismatch between an athlete's action and the rhythm of the music will result in a penalty. Therefore, exploring the consistency of the athlete's action and the rhythm of music is necessary to achieve accurate action quality assessment. Moreover, since the high technical action is usually accompanied by a drastic musical rhythm, music rhythms are natural signals for guiding viewers to focus on important parts of an action. However, previous AQA works \cite{pirsiavash2014assessing, parmar2017learning, xu2019learning, tang2020uncertainty, pan2019action, gao2020asymmetric, zeng2020hybrid, yu2021group, liu2021towards, nagai2021action,nekoui2021eagle, pan2021adaptive, zhang2022semi, zia2018video, doughty2018s, doughty2019pros, li2019manipulation, xu2022likert, bai2022action, parmar2022domain, Xu_2022_CVPR} focus on only visual information and therefore cannot explore such a relation. Thus, we aim to design a multimodal action quality assessment model in this work.}

\ana{It is a fact that the main advantage of multimodal methods is utilizing rich and diverse information from different modalities. However, existing multimodal works of other tasks \cite{owens2018audio, gu2020multi, tian2018audio, hu2021class, wang2022distributed, tao2020audio, liu2019completeness, wu2019dual, lee2020cross, xiao2020audiovisual, gao2020listen, ma2021contrastive, montesinos2022vovit, Xia_2022_CVPR, Jiang_2022_CVPR, badamdorj2021joint, su2020msaf, LiuLWCSQ22, hong2020mini} model modality-specific and multimodal information simultaneously, causing that the information from different modalities inevitably influences each other. In such a way, it is almost impossible to extract pure modality-specific information, and hard to ensure that the network can extract modality-specific information instead of modality-general information. Thus, these methods cannot fully utilize rich and diverse information from different modalities.

To solve the above problem, we propose a multimodal AQA model called \textbf{Progressive Adaptive Multimodal Fusion Network} (PAMFN), which focuses on separately modeling modality-specific information and mixed-modality information. Specifically, our method separately models modality-specific information and mixed-modality information to extract pure modality-specific information via three modality-specific branches and a mixed-modality branch. Thus, our PAMFN is designed as a pyramid architecture with $N$ stages, and the information in the three modality-specific branches is progressively aggregated in the mixed-modality branch during each stage. In addition, to build the bridge between modality-specific branches and the mixed-modality branch, three novel modules are proposed to transfer \lab{information from modality-specific branches} to the mixed-modality branch. 
}

More specifically, to transfer modality-specific information to the mixed-modality branch, a \textbf{Modality-specific Feature Decoder} (MSFD) module is proposed. Since modality-specific features are aggregated in the mixed-modality branch during each stage, the three modality-specific branches and mixed-modality branch contain \laa{certain} overlapping information after the first stage. \laa{Therefore, the MSFD components aims to decode the unseen or neglected information in the mixed-modality branch from modality-specific features.}

Since different parts of an action can be always diverse, we argue that using an invariant multimodal fusion policy may lead to suboptimal results. To solve this issue, an \textbf{Adaptive Fusion Module} (AFM) composed of several \an{fusion networks (FusionNets)} and a \an{policy network (PolicyNet)} is proposed to learn an adaptive multimodal fusion policy. Different FusionNets are used to explore various fusion strategies and the PolicyNet is designed to adaptively select the optimal fusion strategies. In addition, by taking the similarity and diversity among video segments into consideration, we assume that some FusionNets are more general and some are more specific\footnote{If a fusion policy (FusionNet) is suitable for more parts of an action, it is more general. If a fusion policy is suitable for less parts of an action, it is more specific.}; therefore, we propose a novel method called ranked FusionNets in which all FusionNets are ranked and the FusionNet with a higher rank indicates that it is more general. 
We use the Straight-Through Gumbel Estimator \cite{jang2016categorical} to ensure that the decision process is differentiable.

After obtaining the cross-modal features generated by Adaptive Fusion Module, a module called \textbf{Cross-modal Feature Decoder} (CMFD) module is designed to transfer the cross-modal features to the modality-specific branch. Different from the MSFD module, CMFD aims to decode the unseen or neglected information in the mixed-modality branch and modality-specific branches from cross-modal features.

To demonstrate the effectiveness of our method, we conduct extensive experiments on two public action assessment datasets, \ie the Rhythmic Gymnastics dataset \cite{zeng2020hybrid} and the Fis-V dataset \cite{xu2019learning}. Our method achieves state-of-the-art performance on both datasets, showing the advantages of our proposed model. In addition, we evaluate the proposed method on another task, \ie highlight detection, to demonstrate the generalizability of our method. The code for our approach will be released after publication.

\la{Our main contributions can be summarized as follows: 
\begin{itemize}
    \item \laa{We propose a novel multimodal architecture, \lab{called Progressive Adaptive Multimodal Fusion Network (PAMFN)}, for action assessment that separately models modality-specific information and mixed-modality information, and progressively transfers information from modality-specific branches to the mixed-modality branch. To the best of our knowledge, it is the first work to perform action assessment with audio information.}
    \item We propose an Adaptive Fusion Module with a novel ranked FusionNets strategy to learn an adaptive multimodal fusion policy, and use the ST Gumbel Estimator to efficiently train our model.
    \item We propose a Modality-specific Feature Decoder module and a Cross-modal Feature Decoder module to selectively transfer modality-specific information and cross-modal information to the mixed modality branch.
\end{itemize}
}

\section{Related Works}
\subsection{Action Quality Assessment}
Action quality assessment aims to assess the quality of an action and is a fine-grained action understanding task. Based on the length of processed videos, existing works can be divided into two categories: methods for short videos (averagely several seconds) and methods for long videos (averagely several minutes). Many works have focused on AQA for short videos and achieved remarkable progress. 
Wang \etal \cite{wang2021tsa} propose a network to capture rich spatio-temporal contextual information in human motion. \lab{Bai \etal \cite{bai2022action}  propose a temporal parsing transformer to extract fine-grained temporal part-level representations.} On the other hand, quite a few works focus on long videos. 
Zeng \etal \cite{zeng2020hybrid} explore static information and dynamic information for AQA and propose a context-aware attention to learn context information of each segment. Xu \etal \cite{xu2022likert} design a Grade-decoupling Likert Transformer to explore the comprehensive effect of different grades exhibited in the video on the score.

Existing methods explore only visual information in videos ignoring audio information, which is an important cue for assessing the consistency of movement and the rhythm of the music and is a natural signal for guiding us to focus on the important \lab{parts of an action}. Thus, in this work, we propose a multimodal network that leverages RGB, audio and optical flow information to explore modality-specific information and mixed-modality information for AQA. 
\an{Since only long video datasets contain audio among existing AQA datasets, \ie Fis-V dataset \cite{xu2019learning} and Rhythmic Gymnastics dataset \cite{zeng2020hybrid}, our method focuses on long videos and all experiments are conducted on Fis-V dataset and Rhythmic Gymnastics dataset. However, as mentioned in \cite{xu2019learning, zeng2020hybrid}, works \cite{tang2020uncertainty, pan2019action, gao2020asymmetric, yu2021group, liu2021towards, nagai2021action,nekoui2021eagle, pan2021adaptive, zhang2022semi, zia2018video, doughty2018s, doughty2019pros, li2019manipulation, bai2022action, parmar2022domain} designed for short videos are hard to work on long video datasets. Thus, these works will not be compared.}

\subsection{Multimodal Video Understanding}
Multimodal video understanding refers to leveraging different representational modes to understand video, such as RGB frames, optical flows, audio and text. Since the action is usually accompanied with sounds, rich multimodal works\cite{owens2018audio, gu2020multi, tian2018audio, hu2021class, wang2022distributed, tao2020audio, tian2018audio, liu2019completeness, wu2019dual, lee2020cross, xiao2020audiovisual, gao2020listen, ma2021contrastive, montesinos2022vovit, Xia_2022_CVPR, Jiang_2022_CVPR} focus on the interaction of RGB, optical flow and audio information for video understanding. These works are mainly divided into four categories, \ie audio-visual separation and localization, audio-visual recognition, audio-visual representation and audio-visual corresponding learning. In general, the essence of these task are exploring the consistency between audio and video. For instance, \lab{Shi \etal \cite{Shi_2021_ICCV} propose a novel relation model for exploring multimodal multi-action relations in videos, by leveraging both relational graph convolution networks and video multi-modality. Xia \etal \cite{Xia_2022_CVPR} propose a cross-modal time-level and event-level background suppression to better solve the problem of inconsistent audio and visual information within an audiovisual event localization task.} 

\ana{Different from above multimodal works, our method solves the problem that the information from different modalities inevitably influences each other via separately modeling modality-specific information and mixed-modality information, and adaptively selects the optimal fusion policy for each segment conditioned on the input via our Adaptive Fusion Module.}

\begin{figure*}
    \centering
    \includegraphics[width=\linewidth]{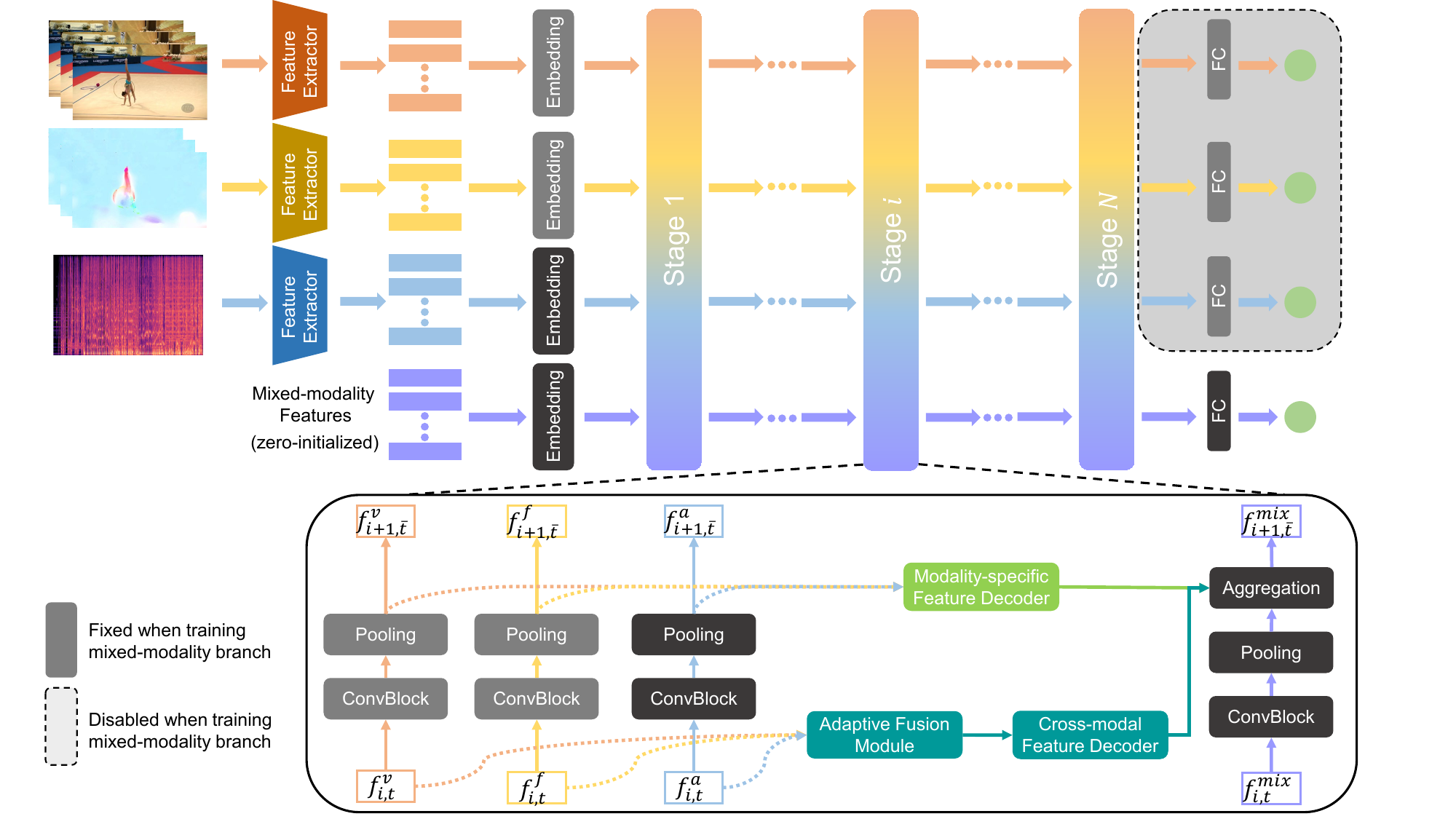}
    \caption{\la{The overall architecture of our proposed PAMFN.} The RGB, optical flow and audio information are fed into three pretrained backbones to extract features respectively. Three modality-specific branches with the same structure are independently pretrained to explore the modality-specific information. Then, a mixed-modality branch progressively aggregates the modality-specific information via a Modality-specific Feature Decoder (MSFD) module and the cross-modal information via an Adaptive Fusion Module (AFM) and a Cross-modal Feature Decoder (CMFD) module. The Adaptive Fusion Module explores an adaptive cross-modal fusion policy. Our network is trained in two phases. We first separately train the modality-specific branches. Then we fix modality-specific branches except the audio branch since the quality of an action is almost impossible to assess using only audio information, and train the mixed-modality branch, MSFD, AFM and CMFD. Note that the initial mixed-modality features are initialized by zeros.}
    \label{fig:framework}
    \vspace{-0.2cm}
\end{figure*}

\section{Approach}
In this section, we introduce the proposed Progressive Adaptive Multimodal Fusion Network (PAMFN). The overall framework is shown in Figure \ref{fig:framework}. First, we describe the preliminaries of our work in Section \ref{method:pre}. Then, we introduce the overview of PAMFN in Section \ref{method:PFA}. Next, we \lab{detail} the three main components of PAMFN, \ie Modality-specific Feature Decoder module, Adaptive Fusion Module and Cross-modal Feature Decoder module, in Sections \ref{method:MSFT}, \ref{method:AFM} and \ref{method:CMFT}, respectively. 

\subsection{Preliminaries}
\label{method:pre}

\noindent{\textbf{Problem Formulation.}} Following previous AQA works \cite{pirsiavash2014assessing,parmar2017learning, zeng2020hybrid, xu2019learning, xu2022likert}, we formulate AQA as a regression problem, where the model observes a video containing a specific action and predicts a non-negative number as the action quality \lab{score}. Similar to \cite{zeng2020hybrid, xu2022likert}, we normalize the labels to the range [0, 1] to ensure stable training:
\begin{equation}
    y^i = \frac {\bar{y} ^ i}  {C},
\end{equation}
where $\bar{y} ^ i$ is the ground truth and $y^i$ is the normalized label for training. $C$ is related to the maximum score of the dataset.

\vspace{0.3cm}

\noindent{\textbf{Feature Extraction.}}
Similar to the common practice in AQA \cite{parmar2017learning, zeng2020hybrid, xu2019learning}, we divide the input video into the non-overlapping video segments. We use the pretrained backbones to extract the features of each video segment, optical flow segment and audio segment. Then, the features of different modalities are fed into different embedding layers and are projected into the same dimension $d$. Formally, given an input video with $T$ segments, we denote the extracted video feature sequence, optical flow feature sequence and audio feature sequence as $\{f^v_t\}^{T}_{t=1}$, $\{f^f_t\}^{T}_{t=1}$ and $\{f^a_t\}^{T}_{t=1}$, respectively. For more details about the extraction process, please refer to Section \ref{expr:details}.

\subsection{Overview of Our Method}
\label{method:PFA}
\laa{To leverage multimodal information for action quality assessment, we propose a novel multimodal architecture. \lab{In this architecture, the} information of different modalities is independently learned and the mixed-modality information is progressively learned from modality-specific information. As shown in Figure \ref{fig:framework}, our PAMFN consists of four branches, \ie a RGB branch, an audio branch, an optical flow branch and a mixed-modality branch. These branches are for learning modality-specific assessment features and learning mixed-modality assessment features below.}

\vspace{0.3cm}

\noindent{\textbf{1) Learning Modality-specific Assessment Features.}} To explore modality-specific information, the modality-specific branches (i.e. the first three branches) except the audio branch are pretrained and fixed since the quality of an action is almost impossible to assess using only audio information (see Section \ref{exp:baseline} for more details). All modality-specific branches have the same structure and consist of $N$ convolution stages and a regression layer. Each stage contains a convolution block and a pooling layer. Each convolution block has the same structure as the residual block in ResNet\cite{he2016deep} except that 1D convolutions are used instead of 2D convolutions. The final regression layer includes a fully-connected layer with a dropout layer and a sigmoid activation function.

\vspace{0.3cm}

\noindent{\textbf{2) Learning Mixed-modality Assessment Features.}} After obtaining the modality-specific information, a mixed-modality branch (i.e. the last branch) is proposed to progressively aggregate the modality-specific information. The mixed-modality branch has same structure with modality-specific branches and the initial mixed-modality features are initialized by zeros. To transfer information from modality-specific branches to the mixed-modality branch, three novel modules are proposed. First, a Modality-specific Feature Decoder module is designed to \lab{selectively} transfer modality-specific information to the mixed-modality branch. Second, an Adaptive Fusion Module is adopted to explore an adaptive multimodal fusion policy for different segments. Third, a Cross-modal Feature Decoder module is used to transfer cross-modal features, generated by Adaptive Fusion Module, to the mixed-modality branch.

\vspace{0.3cm}

\lab{Note that our PAMFN is a pyramid architecture with $N$ stages, and the mixed-modality branch progressively extracts modality-specific information during each stage.}

\begin{figure}
    \centering
    \includegraphics[width=0.7\linewidth]{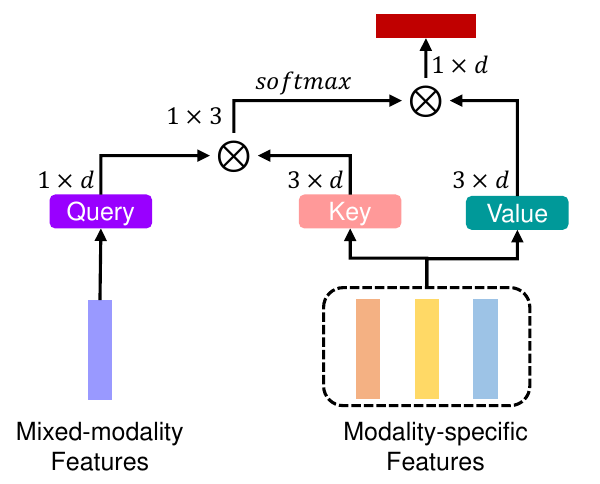}
    \caption{Illustration of the Modality-specific Feature Decoder module. $\otimes$ denotes the matrix multiplication. The shapes of important tensors are shown in the figure. \textit{Query}, \textit{Key} and \textit{Value} are three different linear projections.}
    \label{fig:msfd}
    \vspace{-0.3cm}
\end{figure}


\subsection{Decoding Modality-specific Features}
\label{method:MSFT}
Since the modality-specific features are aggregated in the mixed features during every stage, the mixed features $f^m_{i,t}$ and the modality-specific features, \ie $f^v_{i,t}$, $f^f_{i,t}$ and $f^a_{i,t}$, contain certain the overlapping information in the $i^{th}$ stage. Therefore, a Modality-specific Feature Decoder module is adopted to extract the unseen or neglected information in the mixed-modality branch. Inspired by \cite{vaswani2017attention}, the Modality-specific Feature Decoder module is implemented by a cross-attention. The \textit{query} is the mixed features $f^m_{i,t}$, and the \textit{keys} and \textit{values} are the concatenation of the modality-specific features:
\begin{equation}
    Q_{i,t} = \mathbf{W}^q f^m_{i,t}, \ 
    K_{i,t} = \mathbf{W}^k f^{ms}_{i,t}, \ 
    V_{i,t} = \mathbf{W}^v f^{ms}_{i,t},
\end{equation}
\begin{equation}
    f^{ms}_{i,t} = concatenate(f^v_{i,t}, f^f_{i,t}, f^a_{i,t}),
\end{equation}
where $\{f^v_{i,t}\}_{t=1}^{T_i}$, $\{f^f_{i,t}\}_{t=1}^{T_i}$, $\{f^a_{i,t}\}_{t=1}^{T_i}$ and $\{f^m_{i,t}\}_{t=1}^{T_i}$ have the same dimension $\mathbb{R}^{T_i \times d}$, and $T_i$ represents the length of feature sequences in the $i^{th}$ stage. Then, the cross-attention is formulated as:
\begin{equation}
    \bar{f}^{ms}_{i,t} = softmax(- \frac{Q_{i,t} K_{i,t}^{T}}{\sqrt{d}}) V_{i,t},
\end{equation}
where $\sqrt{d}$ is a scaling factor and $\bar{f}^{ms}_{i,t}$ represents the final modality-specific features that are aggregated in the mixed-modality branch. Taking the negative of $Q_{i,t} K_{i,t}^{T}$ allows us to extract the least similar information between mixed-modality features and modality-specific features. Figure \ref{fig:msfd} shows the details of our Modality-specific Feature Decoder.

\subsection{Learning an Adaptively Fusion Policy}
\label{method:AFM}
\begin{figure}
    \centering
    \includegraphics[width=0.8\linewidth]{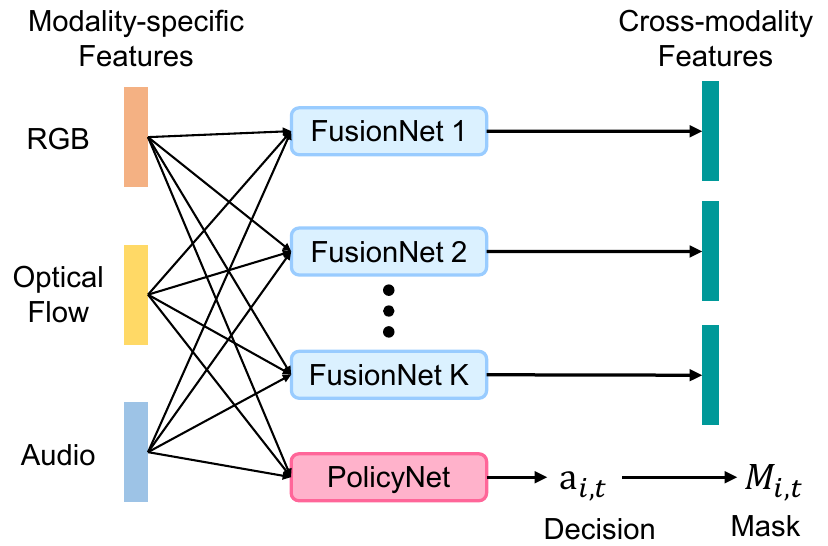}
    \caption{Illustration of the Adaptive Fusion Module. $K$ different FusionNets do not share parameters and explore different fusion policies. The PolicyNet generates an decision $a_{i,t}$ that determines which fusion strategies are enabled. $M_{i,t}$ is a binary mask vector that masks the not enabled cross-modal features. }
    \label{fig:afm}
    \vspace{-0.3cm}
\end{figure}

Since interactions among different modalities are not explicitly modeled in the Modality-specific Feature Decoder module and mixed-modality branch, and using an invariant multimodal fusion policy during different parts of action may lead to suboptimal results, we propose a novel Adaptive Fusion Module to explore an adaptive multimodal fusion policy. The Figure \ref{fig:afm} illustrates an overview of the Adaptive Fusion Module. Our Adaptive Fusion Module consists of $K$ fusion networks called FusionNets and a policy network called PolicyNet, the former for exploring different fusion strategies and the latter for deciding which fusion strategies will be enabled. 

\vspace{0.3cm}

\noindent{\textbf{FusionNet.}} The three modality-specific features are first transformed via three different nonlinear projections. Each FusionNet takes the transformed modality-specific features as inputs and the outputs of the FusionNet, called cross-modal features, are then fed into a convolution block for refinement. The FusionNet is implemented by an attention network that consists of two fully-connected layers followed by a softmax function and the convolution block has the same structure as that in the modality-specific branch. Each attention network assigns weights for different modalities. 

Here, we denote the weights generated by the $k^{th}$ attention network as $\alpha_{i,t,k}^{v}$, $\alpha_{i,t,k}^{f}$ and $\alpha_{i,t,k}^{a}$, and cross-modal features as $f^{cs}_{i,t,k}$. Then, the FusionNet is formulated as:
\begin{equation}
    \bar{f}^{cs}_{i,t,k} = \alpha_{i,t,k}^{v} f_{i-1,t}^{v} + \alpha_{i,t,k}^{f} f_{i-1,t}^{f} + \alpha_{i,t,k}^{a} f_{i-1,t}^{a},
\end{equation}
\begin{equation}
    f^{cs}_{i,t,k} = pool(\Psi_{f}(\bar{f}^{cs}_{i,t,k})),
\end{equation}
where $f_{i-1,t}^{v}$, $f_{i-1,t}^{f}$ and $f_{i-1,t}^{a}$ are modality-specific features in the ${i-1}^{th}$ stage, $\Psi_{f}(\cdot)$ is the convolution block and the pool function is used to ensure that the time dimension is the same as that of the other features in the $i^{th}$ stage. Note that $K$ FusionNets do not share weights, and we use the modality-specific features in the ${i-1}^{th}$ stage instead of those in the $i^{th}$ stage to avoid the influence of the convolution block in the $i^{th}$ stage. Therefore, we obtain $K$ cross-modal features $\{f^{cs}_{i,t,k}\}^{K}_{k=1}$, with each focusing on a different fusion strategy.

\vspace{0.3cm}

\an{
\noindent{\textbf{Ranked FusionNets.}} 
By taking the similarity and diversity among video segments into consideration, we assume that some FusionNets are more general and some are more specific. Then, we proposed a rank mechanism of FusionNets, where the rank of a FusionNet corresponds to its generalizability and the FusionNet with a higher rank indicates it is more general. Specifically, we define that the subscript $k$ ranging from $1$ to $K$ indicates the generality of FusionNet from most to least general. Then, the decision $a_{i,t}=c$ (the output of PolicyNet) represents only the first $c$ FusionNets will be enabled. For example, suppose the decision $a_{i,t}=3$; the top three FusionNets (ranked 1, 2, and 3, respectively) will be enabled, and the FusionNets with lower rank will be disabled. Thus, the FusionNet with a higher rank will be used more frequently and focus on learning the more general fusion strategy. Other definitions of the decision $a_{i,t}$, such as setting $a_{i,t}$ as a one-hot vector (each element indicates the status of corresponding cross-modal features), are discussed in Section \ref{expr:rank fusionnets}.
}

\vspace{0.3cm}

\noindent{\textbf{PolicyNet.}} \an{The PolicyNet, implemented by two fully-connected layers, takes the modality-specific features as inputs and generates the logits $o_{i,t} \in \mathbb{R}^{K}$. As mentioned in Ranked FusionNets, we suppose that the logits $o_{i,t}$ represents the decision $a_{i,t}=c, 1 \le c \le K$, and the generated decision $a_{i,t}=c$ represents the first $c$ cross-modal features $\{f^{cs}_{i,t,k}\}^{c}_{k=1}$ are enabled.} To generate a discrete decision $a_{i,t}$, we first implement the PolicyNet to obtain the action probabilities $P_{i,t} \in \mathbb{R}^{K}$:
\begin{equation}
    P_{i,t} = softmax(\Psi_{p}(f^{ms}_{i-1,t})),
\end{equation}
\begin{equation}
    f^{ms}_{i-1,t} = concatenate(f_{i-1,t}^{v}, f_{i-1,t}^{f}, f_{i-1,t}^{a}),
\end{equation}
where $\Psi_{p}(\cdot)$ is PolicyNet. Then, the $arg \, max$ is used to obtain discrete decision $a_{i,t}$ from the decision probabilities $P_{i,t}$.
Since the $arg$ $max$ is not differentiable, we use the Straight-Through Gumbel Estimator \cite{jang2016categorical} to solve this problem. Specifically, we first use the Gumbel-Max trick \cite{gumbel1954statistical, maddison2014sampling} in the forward process to sample a decision from the decision probabilities $P_{i,t}$:
\begin{equation}
    a_{i,t} = arg \, \underset{k} max  (log P_{i,t,k} + G_{i,t,k} ), 
\end{equation}
where $\{G_{i,t,k}\}_1^K$ are i.i.d samples drawn from the Gumbel(0, 1) distribution $G_{i,t,k} = - log(-log \ U_{i,t,k})$ and $\{U_{i,t,k}\}_1^K$ are i.i.d samples drawn from the uniform distribution $U_{i,t,k} \ \sim \  Uniform(0,1)$. Then, Gumbel-Softmax is used as a continuous, differentiable approximation to differentiate $arg \ max$  in the backward process:
\begin{equation}
    \bar{a}_{i,t,k} = \frac {exp((log \; P_{i,t,k} + G_{i,t,k}) / \tau)} 
    {\sum _{k=1}^{K} exp((log \; P_{i,t,k} + G_{i,t,k}) / \tau)},
\end{equation}
where $\tau$ is the softmax temperature, a non-negative number, and $\bar{a}_{i,t}$ is a continuous approximation of the one-hot encoding representation of ${a}_{i,t}$. \an{$\tau$  controls the concentration level of the distribution. Inspired by CLIP \cite{radford2021learning} and \cite{wu2018unsupervised}, we set the softmax temperature $\tau$ as a learnable parameter.} For more details of Straight-Through Gumbel Estimator, please refer to \cite{jang2016categorical}.

\subsection{Decoding Cross-modal Features}
\label{method:CMFT}

\begin{figure}
    \centering
    \includegraphics[width=0.7\linewidth]{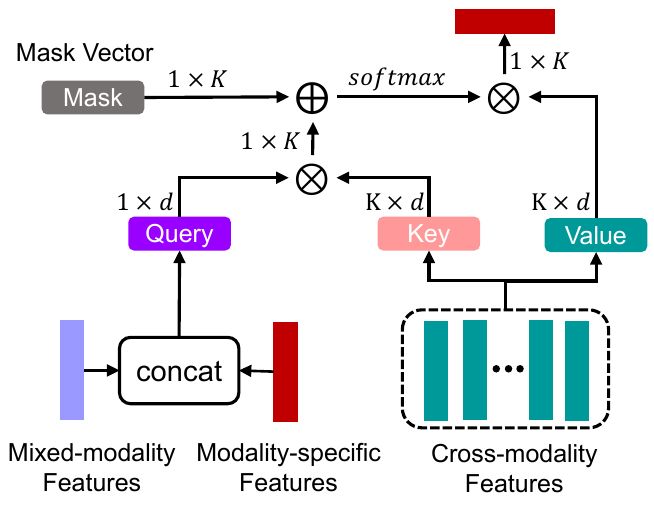}
    \caption{Illustration of the \anb{Cross-modal} Feature Decoder module. $\otimes$ denotes the matrix multiplication. $\oplus$ denotes the element-wise sum. The shapes of important tensors are shown in the figure. \textit{Query}, \textit{Key} and \textit{Value} represent three different linear projections.}
    \label{fig:cmfd}
    \vspace{-0.3cm}
\end{figure}

Similar to the Modality-specific Feature Decoder module, the Cross-modal Feature Decoder module is used to extract unseen or neglected information in the mixed-modality branch. As shown in Figure \ref{fig:cmfd}, the structure of Cross-modal Feature Decoder module is similar to that of the Modality-specific Feature Decoder module, except for the $query$ and an additional mask vector $M_{i,t}=\{m_{i,t,k}\}_{k=1}^K$. The $queries$ are transformed as the concatenation of mixed-modality features and modality-specific features via linear projections. The additional mask vector $M_{i,t}$ masks the disabled cross-modal features and is determined by the decision $a_{i,t}$:
\begin{equation}
        \bar{f}^{cs}_{i,t} = softmax(-\frac{\hat{Q}_{i,t} \hat{K}_{i,t}^{T}}{\sqrt{d}} + M_{i,t}) \hat{V}_{i,t},
\end{equation}
where $\bar{f}^{cs}_{i,t}$ represents the refined cross-modal features. Taking the negative of $\hat{Q}_{i,t} \hat{K}_{i,t}^{T}$ allows us to extract the least similar information from cross-modal features. Since the mask vector $M_{i,t}$ is added before the softmax function, $m_{i,t,k}$ is set as a negative infinite number $\xi$ when the $i^{th}$ cross-features are disabled, or $0$ when the $i^{th}$ cross-features are enabled.

Since we assume that the decision $a_{i,t}=c, 1 \le c \le K$, \lab{indicates that} the first $c$ cross-modal features $\{f^{cs}_{i,t,k}\}^{c}_{k=1}$ are enabled, the mask vector $M_{i,t}$ is generated via a trick in PyTorch \cite{NEURIPS2019_9015}:
\begin{equation}
    M_{i,t} = \bar{M}_{i,t} + \bar{a}_{i,t} - stopgrad(\bar{a}_{i,t}),
\end{equation}
where $\bar{M}_{i,t}$ is a preprocessed mask vector corresponding to the decision $a_{i,t}$. $stopgrad(\cdot)$ is a stop-gradient operation. Therefore, we obtain the mask vector $M_{i,t}$ through a differentiable approach.

\subsection{Model Training}
Since our method focuses on separately modeling modality-specific information and mixed-modality information, we train our network in two stages. In first stage, we train separately each modality-specific branch via a regression layer and the mean-squared error (MSE) loss. In second stage, we fix three modality-specific branches except the audio branch since the quality of an action is almost impossible to assess using only audio information (see Section \ref{exp:baseline} for more details), and train the mixed-modality branch, MSFD, AFM and CMFD via the mean-squared error loss.

\an{
\subsection{Discussion}
\label{sec:disscussion}

\ana{
As mentioned in Section \ref{sec:intro}, different from existing multimodal works \cite{owens2018audio, gu2020multi, tian2018audio, hu2021class, wang2022distributed, tao2020audio, liu2019completeness, wu2019dual, lee2020cross, xiao2020audiovisual, gao2020listen, ma2021contrastive, montesinos2022vovit, Xia_2022_CVPR, Jiang_2022_CVPR, badamdorj2021joint, su2020msaf, LiuLWCSQ22, hong2020mini} which models modality-specific and multimodal information simultaneously, our method separately models modality-specific information and mixed-modality information to extract pure modality-specific information via three modality-specific branches and a mixed-modality branch. Besides, these multimodal works use the same policy to fuse multimodal information for all parts of a video, ignoring diversity among different parts of a long video. By contrast, our method adaptively selects the optimal fusion policy for each segment conditioned on the input via a novel Adaptive Fusion Module. On the other hand, to avoid extracting redundant information from modality-specific branches into the mixed-modality branch in different stages (layers), both MSFD and CMFD aim to extract the unseen or neglected information in the mixed-modality branch, which has not been considered in existing modules. 
}

Additionally, MINI-Net \cite{hong2020mini} also uses several networks named fusion submodules to learn multimodal information, but our FusionNets are constrained by the rank mechanism and PolicyNet to focus on different relations. Besides, with the help of the rank mechanism and PolicyNet, our Adaptive Fusion Module can adaptively decide which FusionNets to enable or disable.
}

\section{Experiments}
\label{sec:expr}
In this section, we first introduce the datasets, evaluation metric and implementation details of our PAMFN. Then, we report the results with state-of-the-art AQA methods and multimodal methods implemented in AQA. Finally, we conduct extensive ablation study experiments and visualize qualitative results to demonstrate the effectiveness of our model.

\subsection{Datasets}

We conduct experiments on two long video AQA datasets, \ie the Fis-V dataset \cite{xu2019learning} and the Rhythmic Gymnastics dataset \cite{zeng2020hybrid}, since only long video datasets contain audio data among the existing AQA datasets.

\vspace{0.3cm}
\noindent{\textbf{Rhythmic Gymnastics (RG) Dataset.}}
The RG dataset contains 1000 videos of four rhythmic gymnastics actions with different apparatuses, \ie ball, clubs, hoop and ribbon. Each video is about 1 minute and 35 seconds with 25 fps and only the duration from the moment of the beginning pose to the moment of the ending pose is preserved in each video. Each video is annotated with three scores, \ie a difficulty sore, an execution score and a total score, given by the referee on the spot. Following the evaluation protocol suggested in \cite{zeng2020hybrid}, we use 200 videos of each gymnastics routine type for training and 50 videos for testing.

\vspace{0.3cm}

\noindent{\textbf{Fis-V Dataset.}} The Fis-V dataset \cite{xu2019learning} contains 500 figure skating videos of the high standard international figure skating competition videos. Each video is about 2 minutes and 50 seconds with about 25 fps and shows the whole performance of only one skater. The irrelevant parts \ie warming up, bowing to the audience and so on, are pruned. Each video is labeled with two labels, namely, Total Element Score (TES) and Total program Component Score (PCS), which are provided by the referee on the spot. Following training-testing split in \cite{xu2019learning}, 400 videos are used for training and the remaining 100 videos are used for testing.

\subsection{Evaluation Metric}

\noindent{\textbf{Spearman’s Rank Correlation (Sp. Corr).}} To compare with previous works \cite{parmar2017learning, xu2019learning, zeng2020hybrid}, Spearman’s rank correlation is used to evaluate our method. Spearman’s rank correlation represents the strength of the relation between the ground-truth labels and the predicted labels and is computed as follows:
\begin{equation}
    \rho = \frac{\sum_{i} (x_i - \bar{x})(y_i - \bar{y})}
    {\sqrt{\sum_{i} (x_i - \bar{x})^2 \sum_{i}(y_i - \bar{y})^2}},
\end{equation}
where $x$ and $y$ represent the rankings of two series. Its value ranges from -1 to 1 and a higher value indicates better results. Following previous works\cite{parmar2017learning, pan2019action, gao2020asymmetric, tang2020uncertainty, wang2021tsa, xu2022likert}, Fisher's z-value is used to compute the average Sp. Corr across actions.

\subsection{Implementation Details}
\label{expr:details}
\noindent{\textbf{Feature Extraction.}} 
As described in Section \ref{method:pre}, we divide RGB frames and optical flows into the non-overlapping segments and each segment contains 32 consecutive frames. To align with the video segments in time, audio is also divided into the same number of segments. Then we use three pretrained models to extract features from RGB frames, optical flows and audio. Following \cite{xu2022likert}, we use the Video Swin Transformer (VST)\cite{zia2018video} pretrained on Kinetics-600 dataset \cite{carreira2017quo} to extract features from RGB frames. For optical flow, we use I3D \cite{carreira2017quo} pretrained on Kinetics-400 dataset\cite{carreira2017quo} to extract features. For audio, we use the Audio Spectrogram Transformer (AST) \cite{gong21b_interspeech} pretrained on the large-scale AudioSet dataset \cite{gemmeke2017audio} to extract audio features from the audio spectrogram. We randomly select 70 consecutive segments on RG and 130 consecutive segments on Fis-V for mini-batch training. If the number of segments is insufficient, we use zero-padding to maintain the same number of segments. All segments are used during testing.

 \begin{table*}[]
\setlength\tabcolsep{2pt}
\centering
\caption{\an{The Spearman's rank correlations of our model compared with the results of state-of-the-art AQA methods on RG and Fis-V datasets. Fisher's z-value is used to compute the average Sp. Corr across actions and the higher values are better. The best results are indicated in bold.}}
\an{
\begin{tabular}{c|c|c|c|ccccc|ccc}
\Xhline{1pt}
\multirow{2}{*}{} & \multirow{2}{*}{\#Params} & \multirow{2}{*}{\makecell[c]{\#Inference \\ Time}} & \multirow{2}{*}{Features} & \multicolumn{5}{c|}{Rhythmic Gymnastics} & \multicolumn{3}{c}{Fis-V} \\ \cline{5-12} 
 & & & & Ball & Clubs & Hoop & Ribbon & \textbf{Avg.} & TES & PCS & \textbf{Avg.} \\ \hline
C3D+SVR\cite{parmar2017learning} & - & - & C3D\cite{tran2015learning} & 0.357 & 0.551  & 0.495 & 0.516 & 0.483 & 0.400  & 0.590  
 & 0.501 \\ \hline
MS-LSTM\cite{xu2019learning} &  1.08M  & 342ms 
    & VST\cite{zia2018video} & 0.621  & 0.661 & 0.670  & 0.695 & 0.663 & 0.660 & 0.809 & 0.744 \\ \hline
ACTION-NET\cite{zeng2020hybrid}  & 3.54M & 2ms 
    & VST\cite{zia2018video}+ResNet\cite{he2016deep} & 0.684 & 0.737 & 0.733 & 0.754 & 0.728 & 0.694 & 0.809 & 0.744 \\ \hline
GDLT\cite{xu2022likert} & 1.84M  & 5ms 
    & VST\cite{zia2018video} & 0.746 & 0.802 & 0.765 & 0.741 & 0.765 & 0.685 & 0.820 & 0.761 \\ \hline
PAMFN(Ours) & 18.06M  & 33ms 
    & VST\cite{zia2018video}+I3D\cite{carreira2017quo}+AST\cite{gong21b_interspeech}
        & \textbf{0.757}	& \textbf{0.825}	& \textbf{0.836}	& \textbf{0.846}	& \textbf{0.819}	& \textbf{0.754}	& \textbf{0.872}	& \textbf{0.822} \\ \Xhline{1pt}
\end{tabular}
}
\label{tab1: aqa}
\end{table*}

\begin{table*}[]
\setlength\tabcolsep{2pt}
\centering
\caption{\an{The Spearman's rank correlations of our model compared with the results of state-of-the-art multimodal methods on RG and Fis-V datasets. Fisher's z-value is used to compute the average Sp. Corr across actions and the higher values are better. The best results are indicated in bold.}}
\an{
\begin{tabular}{c|c|c|c|ccccc|ccc}
\Xhline{1pt}
\multirow{2}{*}{} & \multirow{2}{*}{\#Params} & \multirow{2}{*}{\makecell[c]{\#Inference \\ Time}} & \multirow{2}{*}{Features} & \multicolumn{5}{c|}{Rhythmic Gymnastics} & \multicolumn{3}{c}{Fis-V} \\ \cline{5-12} 
    & & & & Ball & Clubs & Hoop & Ribbo & \textbf{Avg.} & TES & PCS & \textbf{Avg.} \\ \hline
Joint-VA\cite{badamdorj2021joint}  &  1.97M & 4ms & VST\cite{zia2018video}+AST\cite{gong21b_interspeech} 
    & 0.719 & 0.674 & 0.749 & 0.820 & 0.746 & 0.751 & 0.844 & 0.802 \\ \hline
MSAF\cite{su2020msaf} & 5.56M & 10ms & VST\cite{zia2018video}+I3D\cite{carreira2017quo}+AST\cite{gong21b_interspeech} 
    & 0.743 & 0.795 & 0.734 & 0.836 & 0.781 & 0.751 & 0.843 & 0.802 \\\hline
UMT\cite{LiuLWCSQ22} & 3.78M & 5ms & VST\cite{zia2018video}+AST\cite{gong21b_interspeech} 
        & 0.725	& 0.588	& 0.678	& 0.823	& 0.714	& 0.716	& 0.822	& 0.774 \\ \hline
PAMFN(Ours) & 18.06M & 33ms & VST\cite{zia2018video}+I3D\cite{carreira2017quo}+AST\cite{gong21b_interspeech}
        & \textbf{0.757} & \textbf{0.825} & \textbf{0.836}	& \textbf{0.846} & \textbf{0.819} & \textbf{0.754}	& \textbf{0.872}	& \textbf{0.822} \\ \Xhline{1pt}
\end{tabular}
}
\label{tab1: multimodal}
\end{table*}

\vspace{0.3cm}

\noindent{\textbf{Experimental Settings.}}
We implement our PAMFN in three stages ($N = 3$) and set the number of FusionNets $K$ to 10/6 to RG/Fis-V datasets. The feature dimension $d$ is set to 256. The one-head cross-attention is used in MSFD and CMFD. Since our PAMFN is based on pretrained modality-specific branches, our model is trained in two phases. For the modality-specific branches, SGD \cite{lecun1989backpropagation} with a momentum of 0.9 and a weight decay of $10^{-4}$ and a cosine decay learning rate schedule are used to optimize our network. The batch size is 32 and the learning rate is 0.01. We train the modality-specific branches for 250 epochs to get the pretrained modality-specific branches. Then, we train our final model with \an{AdamW \cite{loshchilov2017decoupled} and cosine learning rate decay} after fixing modality-specific branches except for the audio branch. The batch size is 32 and the learning rate is \an{5e-4/8e-4} for RG/Fis-V. Additionally, the learning rate of the regression layer is 0.1 times of previous layers when training the mixed-modality branch. Following \cite{zeng2020hybrid, xu2022likert}, we train different epochs on different datasets for better convergence: \an{400/500/300/500/500/500} for RG(Ball) / RG(Clubs) / RG(Hoop) / RG(Ribbon) / Fis-V(TES) / Fis-V(PCS). \an{Following CLIP\cite{radford2021learning}, the temperature $\tau$ is a learnable parameter and initialized as 10. \anf{Our method is implemented in PyTorch\cite{paszke2019pytorch} and trained on a single RTX 3090Ti GPU.} }

\subsection{Comparison with State-of-the-art Methods}

\an{
To demonstrate the effectiveness of our method, we compare our model with existing AQA methods on RG and Fis-V datasets. Besides, considering that existing AQA methods use only visual information, we also re-implement several multimodal methods used from other tasks for comparison.

\noindent \textbf{- Comparison with AQA methods.}
As shown in Table \ref{tab1: aqa}, our method achieves the best average correlation score on both datasets. Compared with the previous state-of-the-art method GDLT\cite{xu2022likert}, we achieve significant improvement of 0.054 and 0.061 on the average for RG and Fis-V datasets.  \ana{Compared with all previous AQA works using only RGB information, our method fully utilizes multimodal information and explores the consistency of the athlete's action and the rhythm of the music, which helps our method achieve more accurate action quality assessment. \footnotetext{The results of MS-LSTM\cite{xu2019learning} and ACTION-NET\cite{zeng2020hybrid} in Table \ref{tab1: aqa} are from the paper of GDLT\cite{xu2022likert}. MS-LSTM runs slowly since it uses custom LSTM~\cite{hochreiter1997long}.}} 

\noindent \textbf{- Comparison with multimodal methods from other tasks.}
We reimplement three multimodal methods from other tasks, \ie Joint-VA \cite{badamdorj2021joint}, MSAF\cite{su2020msaf} and UMT\cite{LiuLWCSQ22}. As shown in Table \ref{tab1: multimodal}, our method outperforms these multimodal methods on both RG and Fis-V datasets. Due to the benefits of separately modeling modality-specific information and mixed-modality information and adaptive fusion policy, our method achieves more accurate action quality assessment. Note that Joint-VA and UMT are not designed for AQA and UMT has some task-specific modules or losses, so it is not surprising that MSAF  outperforms Joint-VA and UMT. 

According to the results in Table \ref{tab1: aqa} and Table \ref{tab1: multimodal}, we find that although these multimodal methods are not designed for AQA, these methods outperform all existing AQA methods with the help of audio information, demonstrating the benefits of audio information mentioned in the introduction. Besides, although our method takes more time in inference than existing AQA methods, it is still tolerable since it costs only 33ms which is fast enough, and the primary time cost is for the feature extraction but not the assessing network.  

\ana{Additionally, since the used feature extractors may seem slightly suboptimal, we conduct experiments with stronger feature extractors to demonstrate the effectiveness of our method. Specifically, UNMT \cite{li2023unmasked} pretrained on Kinectics-600 dataset and MAST \cite{zhu2023multiscale} ptretrained on AudioSet are used as RGB and audio feature extractors. As shown in Table \ref{tab1:strong}, with the help of stronger features extracted by UNML and MAST, the performance of all methods boosts and our method achieves 0.835 and 0.832 on the average for RG and Fis-V datasets.}
}

\begin{table*}[]
\setlength\tabcolsep{2pt}
\centering
\caption{\an{The Spearman's rank correlations when using strong features extracted by UNMT\cite{li2023unmasked} and MAST\cite{zhu2023multiscale} on RG and Fis-V datasets. Fisher's z-value is used to compute the average Sp. Corr across actions and the higher values are better. The best results are indicated in bold.}}
\an{
\begin{tabular}{c|c|ccccc|ccc}

\Xhline{1pt}
\multirow{2}{*}{} & \multirow{2}{*}{Features} & \multicolumn{5}{c|}{Rhythmic Gymnastics} & \multicolumn{3}{c}{Fis-V} \\ \cline{3-10} 
 & & Ball & Clubs & Hoop & Ribbon & \textbf{Avg.} & TES & PCS & \textbf{Avg.} \\ \hline
GDLT\cite{xu2022likert} & UNMT\cite{li2023unmasked} & 0.785	& 0.776	& 0.768	& 0.776	& 0.776 & 0.710 & 0.823 & 0.773 \\ \hline
Joint-VA\cite{badamdorj2021joint}  & UNMT\cite{li2023unmasked}+MAST\cite{zhu2023multiscale} &
     0.735 & 0.648 & 0.811 & 0.826 & 0.763 & 0.768 & 0.849 & 0.812\\ \hline
MSAF\cite{su2020msaf}& UNMT\cite{li2023unmasked}+I3D\cite{carreira2017quo}+MAST\cite{zhu2023multiscale}
        & 0.780 & 0.740 & 0.802 & 0.842 & 0.794 & 0.779 & 0.855 & 0.821 \\\hline
UMT\cite{LiuLWCSQ22} & UNMT\cite{li2023unmasked}+MAST\cite{zhu2023multiscale}
        & 0.736&0.651&0.801&0.7998&0.753&0.732&0.810&0.774 \\\hline
PAMFN(Ours) & UNMT\cite{li2023unmasked}+I3D\cite{carreira2017quo}+MAST\cite{zhu2023multiscale}
        & \textbf{0.822} & \textbf{0.813} & \textbf{0.853} & \textbf{0.848}	& \textbf{0.835} & \textbf{0.791} & \textbf{0.865} & \textbf{0.832} \\ \Xhline{1pt}

\end{tabular}
}
\label{tab1:strong}
\end{table*}

\begin{table*}[]
\centering
\caption{The \an{Spearman's rank correlations} of our model compared with the results of the unimodality methods and multimodal methods using Weighted Fusion on RG and Fis-V datasets. }
\label{tab2}
\an{
\begin{tabular}{c|ccc|ccccc|ccc}
\hline
\multirow{2}{*}{} & \multicolumn{3}{c|}{Modalities} & \multicolumn{5}{c|}{Rhythmic Gymnastics} & \multicolumn{3}{c}{Fis-V} \\ \cline{2-12} 
        & RGB   & Flow  & Audio     & Ball      & Clubs     & Hoop      & Ribbon    & \textbf{Avg.}  & TES   & PCS   & \textbf{Avg.} \\ \hline
\multirow{3}{*}{Unimodality Methods} 
    & $\checkmark$ &              &              & 0.636 & 0.720 & 0.769 & 0.708 & 0.711 & 0.665 & 0.823 & 0.755 \\
    &              & $\checkmark$ &              & 0.536 & 0.674 & 0.684 & 0.716 & 0.657 & 0.606 & 0.772 & 0.698 \\
    &              &              & $\checkmark$ & 0.286 & 0.297 & 0.423 & 0.254 & 0.317 & 0.517 & 0.628 & 0.575 \\ \hline
\multirow{3}{*}{\begin{tabular}[c]{@{}l@{}} Multimodal Methods \\using Weighted Fusion \cite{panda2021adamml}\end{tabular}}                               
    & $\checkmark$ & $\checkmark$ &              & 0.679 & 0.736 & 0.754 & 0.735 & 0.727 & 0.643 & 0.838 & 0.757 \\
    & $\checkmark$ &              & $\checkmark$ & 0.601 & 0.623 & 0.746 & 0.749 & 0.686 & 0.735 & 0.825 & 0.784 \\
    &              & $\checkmark$ & $\checkmark$ & 0.514 & 0.493 & 0.561 & 0.629 & 0.552 & 0.637 & 0.782 & 0.717 \\
    & $\checkmark$ & $\checkmark$ & $\checkmark$ & 0.613 & 0.647 & 0.763 & 0.763 & 0.703 & 0.733 & 0.848 & 0.798 \\ \hline
\multirow{4}{*}{Ours}          
    & $\checkmark$ & $\checkmark$ &              &  0.750 & 0.788 & 0.791 & 0.824 & 0.790 & 0.742 & 0.867 & 0.814\\
    & $\checkmark$ &              & $\checkmark$ &  0.749 & 0.757 & \textbf{0.856} & 0.769 & 0.787 & 0.736 & 0.863 & 0.809\\
    &              & $\checkmark$ & $\checkmark$ &  0.693 & 0.665 & 0.795 & 0.795 & 0.743 & 0.689 & 0.813 & 0.758\\ \cline{2-12} 
    & $\checkmark$ & $\checkmark$ & $\checkmark$ & \textbf{0.757} & \textbf{0.825} & 0.836	& \textbf{0.846} & \textbf{0.819} & \textbf{0.754}	& \textbf{0.872}	& \textbf{0.822} \\ \hline
\end{tabular}
}
\end{table*}

\subsection{Comparison with Baselines}
\label{exp:baseline}
Since all existing AQA methods use only the RGB information, we compare our method with following multimodal baselines:
\begin{itemize}
    \item Unimodality methods. We compare our method with modality-specific baselines in which we train three different models using RGB, optical flow and audio, respectively. The structure of each modality-specific model is the same to the modality-specific branch of our methods.
    \item Multimodal methods using Weighted Fusion. We compare our method with four joint learning multimodal baselines in which each model uses different combinations of the three modalities. A simple method, called Weighted Fusion\cite{panda2021adamml}, is used to fusion different modalities via late fusion with learnable weights in each multimodal baseline. And a softmax function is used to guarantee that the weight of each modality is positive. Note that these joint learning multimodal baselines are trained in one stage.
    \item We compare our method with different combinations of three modalities by removing one modality-specific branch from our model. Note that these baselines are trained in two stages, similar to our method.
\end{itemize}

Table \ref{tab2} shows the results of the above baseline comparisons on RG and Fis-V. Although the modality-specific branch of PAMFN is a simple network with three convolution blocks, the unimodality method using only RGB achieves a competitive performance (0.755) when compared with GDLT \cite{xu2022likert} (0.761) on Fis-V. Since AQA focuses on human actions, the unimodality method using only audio naturally obtains poor results, which is why we finetune the audio branch instead of fixing it. Additionally, it's surprise that the unimodality method using only audio achieves a performance of 0.575, which reveals that the action quality on skating has a strong correlation with the background music. 
Most multimodal methods using Weighted Fusion achieve worse results than unimodality methods on RG dataset but obtain conflicting results on Fis-V, which is most likely because the action quality on skating has a strong correlation with the background music. As shown in Table \ref{tab2}, our method clearly outperforms all multimodal methods using Weighted Fusion. In summary, multimodal information can improve the performance and our model effectively uses the multimodal information. Note that the mixed-modality branch is used in all multimodal methods, and experiments without the mixed-modality branch are described in Section \ref{EFS}.

\subsection{Ablation Study}

\begin{table*}[]
\centering
\caption{Evaluation of Fusion Strategies.}
\label{tab3}
\an{
\begin{tabular}{c|c|ccccc|ccc}
\hline
\multirow{2}{*}{} & \multirow{2}{*}{Fusion} & \multicolumn{5}{c|}{Rhythmic Gymnastics} & \multicolumn{3}{c}{Fis-V} \\ \cline{3-10} 
    & & Ball & Clubs & Hoop & Ribbon & \textbf{Avg.} & TES & PCS & \textbf{Avg.} \\ \hline
\multirow{5}{*}{One-stage Training}     
    & AVG       & 0.596 & 0.644 & 0.746 & 0.778 & 0.698     & 0.715 & 0.840 & 0.785     \\
    & CAT       & 0.512 & 0.585 & 0.674 & 0.694 & 0.621     & 0.691 & 0.806 & 0.754     \\
    & Weighted\cite{panda2021adamml}  & 0.613 & 0.647 & 0.763 & 0.763 & 0.703     & 0.733 & 0.848 & 0.798     \\
    & Attention & 0.566 & 0.639 & 0.718 & 0.830 & 0.703     & 0.649 & 0.814 & 0.743     \\
    & Ours      & 0.741 & 0.728 & 0.819 & \textbf{0.855} & 0.792     & 0.735 & 0.853 & 0.802     \\ \hline
\multirow{5}{*}{Two-stage Training} 
    & AVG       & 0.637 & 0.747 & 0.781 & 0.771 & 0.739     & 0.720 & 0.857 & 0.799     \\
    & CAT       & 0.665 & 0.762 & 0.805 & 0.817 & 0.768     & 0.696 & 0.850 & 0.785     \\
    & Weighted\cite{panda2021adamml}  & 0.638 & 0.752 & 0.781 & 0.776 & 0.742     & 0.721 & 0.860 & 0.801     \\
    & Attention & 0.641 & 0.744 & 0.778 & 0.773 & 0.738     & 0.706 & 0.847 & 0.787     \\
    & Ours      & \textbf{0.757} & \textbf{0.825}	& \textbf{0.836} & 0.846 & \textbf{0.819} & \textbf{0.754} & \textbf{0.872} & \textbf{0.822}  \\ \hline
\end{tabular}
}
\vspace{-0.3cm}
\end{table*}

\begin{table*}[]
\centering
\caption{Evaluation of Cross-modal Feature Decoder (CMFD) module and Modality-specific Feature Decoder (MSFD) module. $\rightarrow$ denotes replacing a module with Weighted Fusion.}
\label{tab4}
\an{
\begin{tabular}{c|ccccc|ccc}
\hline
\multirow{2}{*}{Fusion} & \multicolumn{5}{c|}{Rhythmic Gymnastics} & \multicolumn{3}{c}{Fis-V} \\ \cline{2-9} 
 & Ball & Clubs & Hoop & Ribbon & \textbf{Avg.} & TES & PCS & \textbf{Avg.} \\ \hline
w/o CMFD & 0.672 & 0.789 & 0.769 & 0.775 & 0.755 & 0.733 & 0.858 & 0.804 \\
w/o MSFD & 0.700 & 0.814 & 0.833 & 0.839 & 0.802 & 0.734 & 0.871 & 0.813 \\
Weighted Fusion\cite{panda2021adamml} $\rightarrow$ CMFD & 0.670 & 0.751 & 0.816 & 0.796 & 0.764 & 0.724 & 0.862 & 0.804 \\
Weighted Fusion\cite{panda2021adamml} $\rightarrow$ MSFD & \textbf{0.773} & 0.782 & 0.810 & 0.836 & 0.802 & 0.726 & 0.862 & 0.804 \\
Weighted Fusion\cite{panda2021adamml} $\rightarrow$ MSDF and CMFD & 0.629 & 0.804 & 0.819 & 0.816 & 0.777 & 0.712 & 0.858 & 0.796 \\ \hline
Our Full Model & 0.757 & \textbf{0.825} & \textbf{0.836} & \textbf{0.846} & \textbf{0.819} & \textbf{0.754} & \textbf{0.872} & \textbf{0.822} \\ \hline
\end{tabular}
}
\end{table*}

\subsubsection{Evaluation of Fusion Strategies}
\label{EFS}

To demonstrate the effectiveness of our fusion strategy, we first compare with four common late fusion strategies:
\begin{itemize}
    \item AVG: We simply calculate the average of three modality features.
    \item CAT: We concatenate three modality features and use an additional fully connected (FC) layer to reduce the dimension.
    \item Weighted\an{\cite{panda2021adamml}}: Learnable weights and a softmax function are used to fuse three modality features as described above.
    \item Attention: An FC layer takes the three modality features as inputs and generates weights for each modality. Then the generated weights are fed into a softmax function and the fused features is obtained by computing a weighted sum.
\end{itemize}

Since our method separately models modality-specific information and cross-modal information, we train our mixed-modality branch based on pretrained modality-specific branches. To further validate the superiority of our method and the effectiveness of two-stage training strategy, we train the above baselines with two-stage training and one-stage training. For one-stage training, fusion strategies are used to combine the outputs of the last convolution block, and two FC layers used as a regressor take the fused features as inputs.

The results are shown in Table \ref{tab3}. Our method with two-stage training achieves the best average performance on RG and Fis-V, and outperforms other fusion strategies with a large margin on average. We observe that the models trained in two stages achieve better performance than the models trained in one stage. Remarkably, the models trained in one stage achieve worse average performance than the modality-specific model only using RGB on RG dataset, and we get an opposite conclusion on Fis-V. The weighted fusion performs better than the compared with other fusion strategies. Additionally, the average performance of our method decreases by \an{0.027 and 0.02 on RG and Fis-V} when trained in one stage instead of in two stages, most likely because our method is designed based on the pretrained modality-specific branches and the modality-specific branches have difficulty extracting high-quality modality-specific information when trained in one stage. 

\vspace{0.3cm}

\subsubsection{Evaluation of the CMFD and MSFD}

To demonstrate the effectiveness of our proposed Cross-modal Feature Decoder module and Modality-specific Feature Decoder module, we try to remove Cross-modal Feature Decoder and Modality-specific Feature Decoder from the full model. Additionally, we try to replace Cross-modal Feature Decoder and Modality-specific Feature Decoder with Weighted Fusion to show the superiority of our proposed components. Here, we choose Weighted Fusion because Weighted Fusion achieves the best performance among four common fusion strategies.

The results are shown in Table \ref{tab4} and $\rightarrow$ denotes replacing a module with Weighted Fusion. Removing the Cross-modal Feature Decoder or Modality-specific Feature Decoder causes an average performance drop of \an{0.064/0.018 and 0.017/0.009} on RG/Fis-V respectively, which shows that modality-specific information is more important than cross-modal information for action quality assessment. Replacing the Cross-modal Feature Decoder or Modality-specific Feature Decoder with Weighted Fusion alleviates the performance drop when compared with removing one of decoders. However, completely replacing two decoders with Weighted Fusion also leads to a performance drop. These results demonstrate the effectiveness of our Cross-modal Feature Decoder and Modality-specific Feature Decoder.

\vspace{0.3cm}

\begin{figure*}[]
	\centering
		\subfloat[Stage 1]{\includegraphics[width=.3\textwidth]{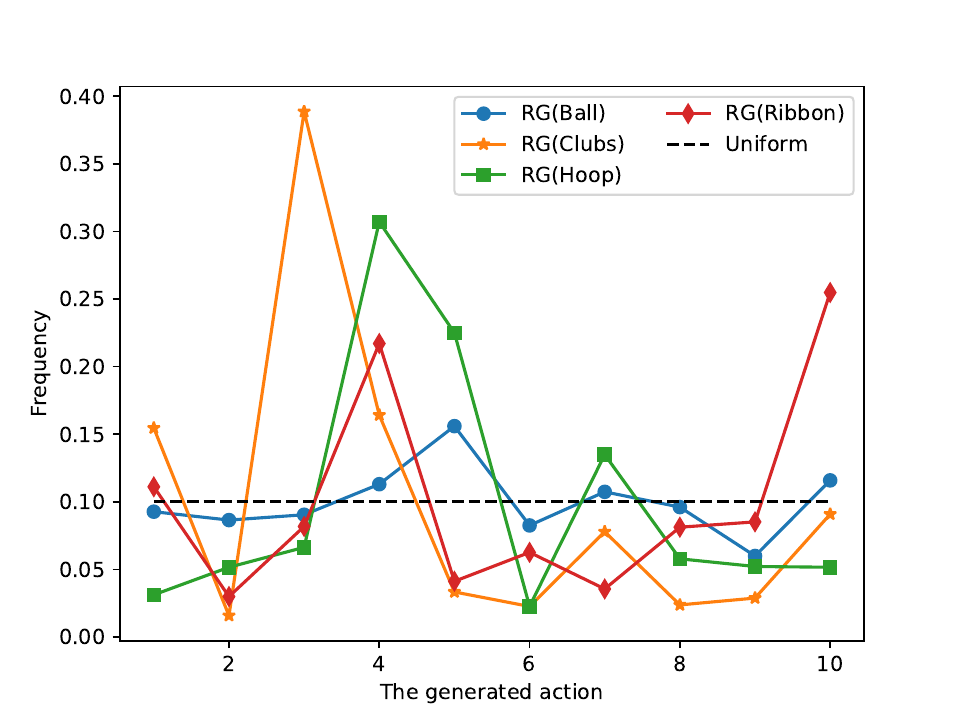}%
		\label{fig:freq_1}}
	\hfil
	    \subfloat[Stage 2]{\includegraphics[width=.3\textwidth]{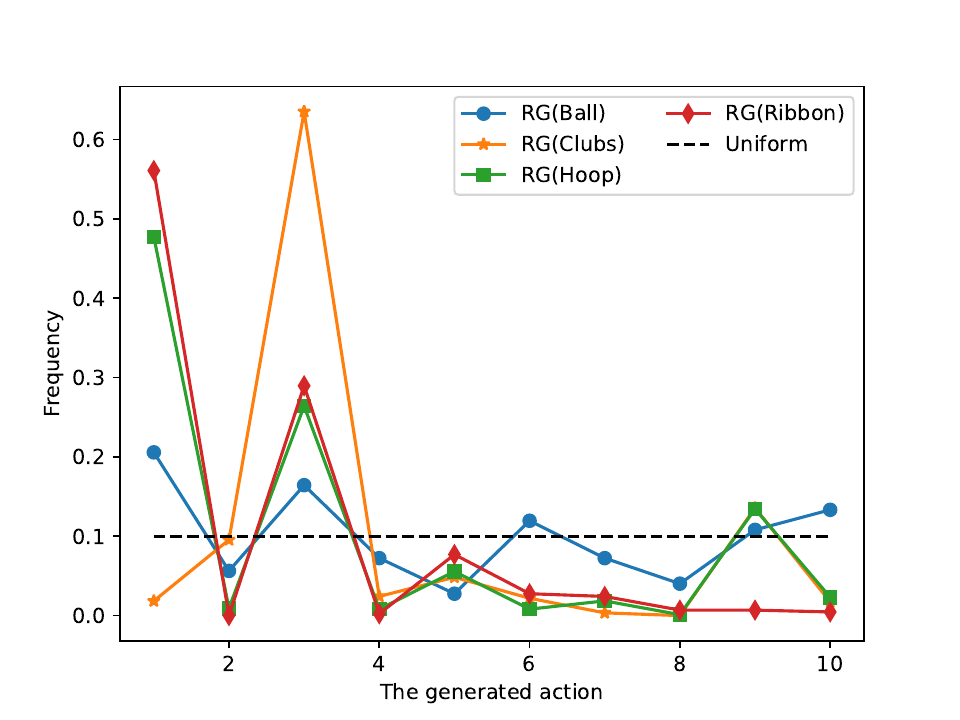}%
		\label{fig:freq_2}}
	\hfil
	    \subfloat[Stage 3]{\includegraphics[width=.3\textwidth]{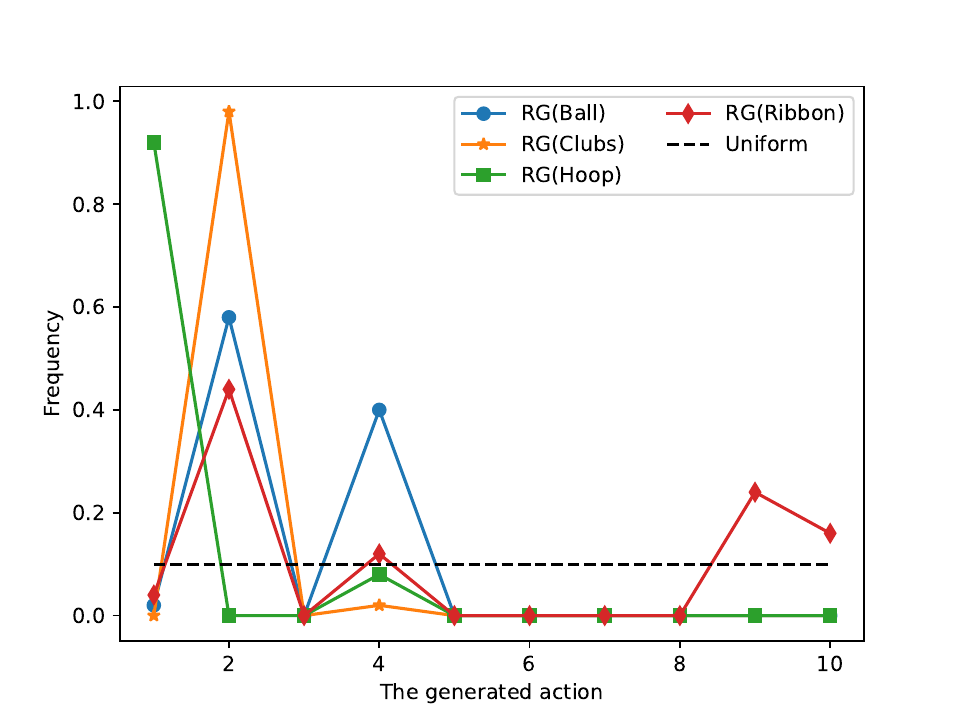}%
		\label{fig:freq_3}}
	\caption{\an{The frequency of the generated decisions at three stages on RG dataset. The line labeled with Uniform denotes the frequency of decisions under a uniform distribution. Best viewed in color.}}
	\label{fig:action_freq}
	\vspace{-0.3cm}
\end{figure*}

\begin{table*}[]
\centering
\caption{Evaluation of different FusionNets on RG and Fis-V datasets. }
\label{tab5}
\an{
\begin{tabular}{c|ccccc|ccc}
\hline
\multirow{2}{*}{} & \multicolumn{5}{c|}{Rhythmic Gymnastics} & \multicolumn{3}{c}{Fis-V} \\ \cline{2-9} 
    & Ball & Clubs & Hoop & Ribbon & \textbf{Avg.} & TES & PCS & \textbf{Avg.} \\ \hline
Unranked FusionNets      & 0.689 & 0.822 & 0.789 & 0.832 & 0.789 & 0.730 & 0.868 & 0.810 \\
Free FusionNets          & 0.755 & 0.793 & 0.835 & 0.843 & 0.809 & 0.733 & 0.870 & 0.812 \\ \hline
Ranked FusionNets (Ours) & \textbf{0.757} & \textbf{0.825} & \textbf{0.836} & \textbf{0.846} & \textbf{0.819} & \textbf{0.754} & \textbf{0.872} & \textbf{0.822} \\ \hline
\end{tabular}
}
\vspace{-0.1cm}
\end{table*}

\begin{table*}[]
\centering
\caption{Evaluation of fusing modality information at different stages on RG and Fis-V datasets.}
\label{tab6}
\an{
\begin{tabular}{c|ccccc|ccc}
\hline
\multirow{2}{*}{Fusion Stage} & \multicolumn{5}{c|}{Rhythmic Gymnastics} & \multicolumn{3}{c}{Fis-V} \\ \cline{2-9} 
    & Ball & Clubs & Hoop & Ribbon & \textbf{Avg.} & TES & PCS & \textbf{Avg.} \\ \hline
First Stage       & 0.733 & 0.785 & 0.795 & 0.824 & 0.786 & 0.753 & 0.863 & 0.815 \\
First Two Stages  & 0.730 & 0.786 & \textbf{0.836} & \textbf{0.864} & 0.810 & \textbf{0.755} & \textbf{0.873} & \textbf{0.823} \\ 
All Stages (Ours) & \textbf{0.757} & \textbf{0.825} & \textbf{0.836} & 0.846 & \textbf{0.819} & 0.754 & 0.872 & 0.822 \\ \hline
\end{tabular}
}
\end{table*}

\subsubsection{Evaluation of Ranked FusionNets}
\label{expr:rank fusionnets}
To demonstrate the effectiveness of our proposed ranked FusionNets, we compare our ranked FusioNets with unranked FuionNets. To implement unranked FusionNets, we define that the decision $a_{i,t} \in R^{K}$, generated by the PolicyNet, is a one-hot vector and each element denotes the corresponding FusionNet is enabled or disabled. We also try to enable all FusionNets called Free FusionNets and remove the PolicyNet since all FusionNets is enabled. Note that the numbers of FusionNets $K$ remains the same in each model.

As shown in Table \ref{tab5}, replacing ranked FusionNets with unranked FusionNets or Free FusionNets causes an average performance drop of \an{0.03/0.012 and 0.01/0.009} on RG/Fis-V, respectively. Although unranked FusionNets has more freedom to select the optimal combination of FusionNets than ranked FusionNets, the model has difficulty converging to the optimal solution and achieves poor results. Similarly, most information, \ie all FusionNets, is used in Free FusionNets, but this method also obtains poor results, especially on RG. Different from unranked FusionNets and free FusionNets, since we assume that the FusionNet with a higher rank indicates that it is more general and the first $c$ FusionNets are enabled when given an decision $a_{i,t} = c$, the model converges more easily than the other two models.

To verify that our ranked FusionNets does not converge to a trivial solution, \ie the generated decisions are almost the same and only a few FusionNets are used, we visualize the generated decisions on the RG test set. The frequencies of the generated decisions in three stages are shown in Figure \ref{fig:action_freq} and the line labeled Uniform represents the frequency of decisions according to a uniform distribution. As shown in Figure \ref{fig:action_freq}, our ranked FusionNets does not converge to a trivial solution \an{and our method really adopts different fusion policies for different parts of an action. We notice that although we adopt the same number of FusionNets across all actions for simplicity, our method can learn to disable redundant FusionNets, as shown in Figure \ref{fig:action_freq}-c.} Remarkably, the frequency curves of different actions have similar trends, most likely because these actions are all rhythmic gymnastics actions, with only the apparatus differing. 

\vspace{0.3cm}

\subsubsection{Evaluation of the Progressive Fusion Strategy}

Our method is based on a progressive fusion strategy which means that the mixed-modality branch extracts information from the modality-specific branches in different stages. To demonstrate the effectiveness of our progressive fusion strategy, we try to only fuse modality information at specific layers. Specifically, we evaluate our method with fusion in only the first stage or in the first two stages. 
The results are shown in Table \ref{tab6}. We can observe that fusion during the first stage achieves the worst performance of \an{0.786} on RG and the performance improves as the number of fusion stages increases. In addition, fusion during the first stage or first two stages leads to a slight performance drop on Fis-V.

\vspace{0.3cm}

\subsubsection{Evaluation of the Number of FusionNets $K$}

\begin{figure}
    \vspace{-0.5cm}
    \centering
    \includegraphics[width=0.8\linewidth]{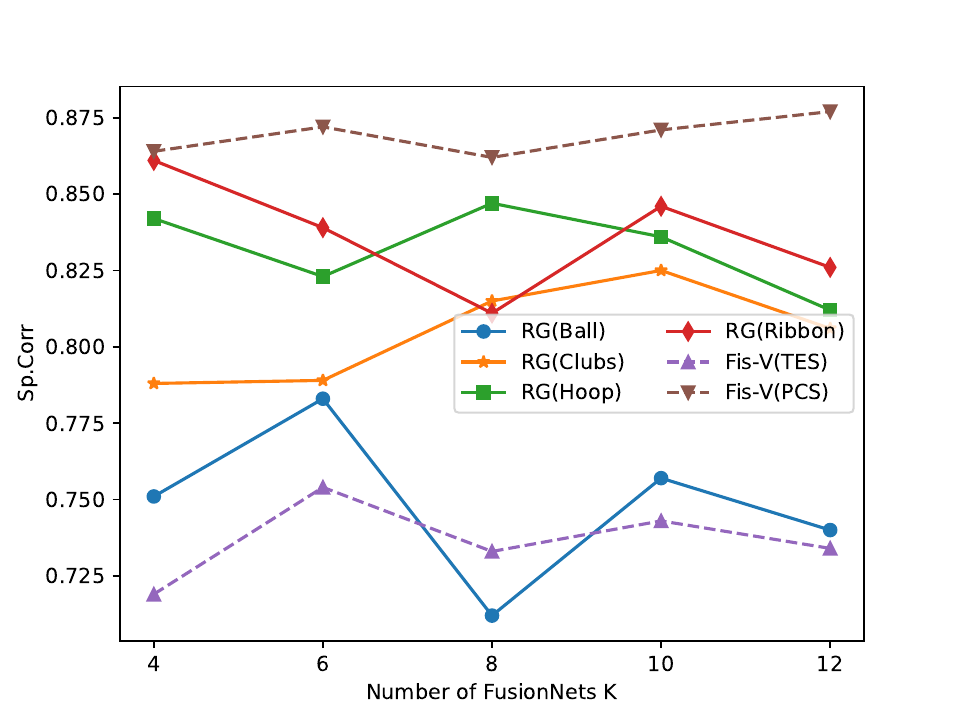}
    \caption{\an{Evaluation of the number of FusionNets.}}
    \label{fig:ablation_study}
    \vspace{-0.2cm}
\end{figure}

We implement our method with different numbers of FusionNets $K$ to evaluate the impact of the number of FusionNets on model performance. As shown in Figure \ref{fig:ablation_study}, different action types has different optimal $K$ values, and $K = 10/6$ achieves the best average performance on RG/Fis-V dataset. The performance on RG/Fis-V dataset decreases slightly when the number of FusionNets is larger than 10. In addition, we observe clear changes in the performance as the number of FusionNets ranges from 4 to 12 on RG and the number of FusionNets is more important for RG than Fis-V.

\begin{table*}[]
\centering
\caption{The \an{mean average precision} of our model compared with the results of state-of-the-art methods on YouTube Highlights dataset. The best results are indicated in bold and the second-best results are underlined. }
\label{tab:hd}
\begin{tabular}{c|c|ccccccc}
\hline
Method   & Modality & Dog            & Gymnastics     & Parkour        & Skating         & Skiing         & Surfing        & \textbf{Average}        \\ \hline
RRAE\cite{yang2015unsupervised} & RGB & \an{0.490} & \an{0.350} & \an{0.500} & \an{0.250} & \an{0.220} & \an{0.490} & \an{0.383} \\
GIFs\cite{gygli2016video2gif}   & RGB & 0.308 & 0.335 & 0.540 & 0.554 & 0.328 & 0.541 & 0.464  \\
LSVM\cite{sun2014ranking}       & RGB & \an{0.600} & \an{0.410} & \an{0.610} & \an{0.620} & \an{0.360} & \an{0.610} & \an{0.536} \\
CLA\cite{xiong2019less}         & RGB & 0.502 & 0.217 & 0.309 & 0.505 & 0.379 & 0.584 & 0.416 \\
LM\cite{xiong2019less}          & RGB & 0.579 & 0.417 & 0.670 & 0.578 & 0.486 & 0.651 & 0.564 \\
Mini-Net\cite{hong2020mini}     & RGB+Audio & \an{0.582} & \an{0.617} & \an{0.702} & \an{0.722} & \an{0.587} & \an{0.651} & \an{0.644} \\
Trail.\cite{wang2020learning}   & RGB & 0.633 & \textbf{0.825} & 0.623 & 0.529 & \textbf{0.745} & 0.793 & 0.691 \\
DL-VHD\cite{badamdorj2021joint}  & RGB & 0.708 & 0.532 & 0.772 & \underline{0.725} &0.661 & 0.762 & 0.693 \\
Joint-VA\cite{badamdorj2021joint}       & RGB+Audio & 0.645 & 0.719 & 0.808 & \an{0.620} & \underline{0.732} & 0.783 & 0.718 \\ 
\an{PLD\cite{Wei_2022_CVPR}}     & \an{RGB} & \an{\textbf{0.749}} & \an{0.702} & \an{0.779} & \an{0.575} & \an{0.707} & \an{0.790} & \an{0.730} \\
\an{CO-AV\cite{LiZYLLHY22}}   & \an{RGB+Audio} & \an{0.609} & \an{0.660} & \an{\textbf{0.890}} & \an{\textbf{0.741}} & \an{0.690} & \an{0.811} & \an{\underline{0.747}} \\
\an{UMT\cite{LiuLWCSQ22}}     & \an{RGB+Audio} & \an{0.659} & \an{\underline{0.752}} & \an{0.816} & \an{0.718} & \an{0.723} & \an{\textbf{0.827}} & \an{\textbf{0.749}} \\ \hline
\multirow{4}{*}{PAMFN(Ours)}     
     & RGB          & \an{0.652} & \an{0.677} & \an{0.591} & \an{0.629} & \an{0.710} & \an{\underline{0.825}} & \an{0.681} \\
     & Flow         & \an{0.626} & \an{0.657} & \an{0.662} & \an{0.420} & \an{0.672} & \an{0.684} & \an{0.620} \\
     & Audio        & \an{0.502} & \an{0.639} & \an{0.720} & \an{0.467} & \an{0.655} & \an{0.737} & \an{0.620} \\
     & RGB+Audio+Flow & \an{\underline{0.720}} & \an{0.690} & \an{\underline{0.822}} & \an{0.722} & \an{0.720} & \an{0.810} & \an{\underline{0.747}} \\ \hline
\end{tabular}
\end{table*}

\subsection{Extension to Highlight Detection}

\la{Although our method is designed for action quality assessment, the architecture of our method is general and \laa{can be easily used for} other tasks. To demonstrate the generalizability of our method, we evaluate our method on the highlight detection task, which is a task of detecting interesting moments called ``highlights", within videos. To apply our method, we remove all pooling layers and replace the regression layer with a classification head. We conduct experiments on YouTube Highlights dataset\cite{sun2014ranking} which is a commonly used dataset for highlight detection. 

\vspace{0.3cm}

\noindent{\textbf{Dataset.}} YouTube Highlights dataset\cite{sun2014ranking} contains six topic categories, \ie dog, gymnastics, parkour, skating, skiing, and surfing, and each topic contains about 100 videos. The annotations are segment-level and indicate whether a segment is a highlight segment. We follow the training-test split \an{\cite{LiuLWCSQ22}} to evaluate our model. Moreover, we train a highlight detector for each topic, following existing works.

\begin{figure*}
    \centering
    \includegraphics[width=\linewidth]{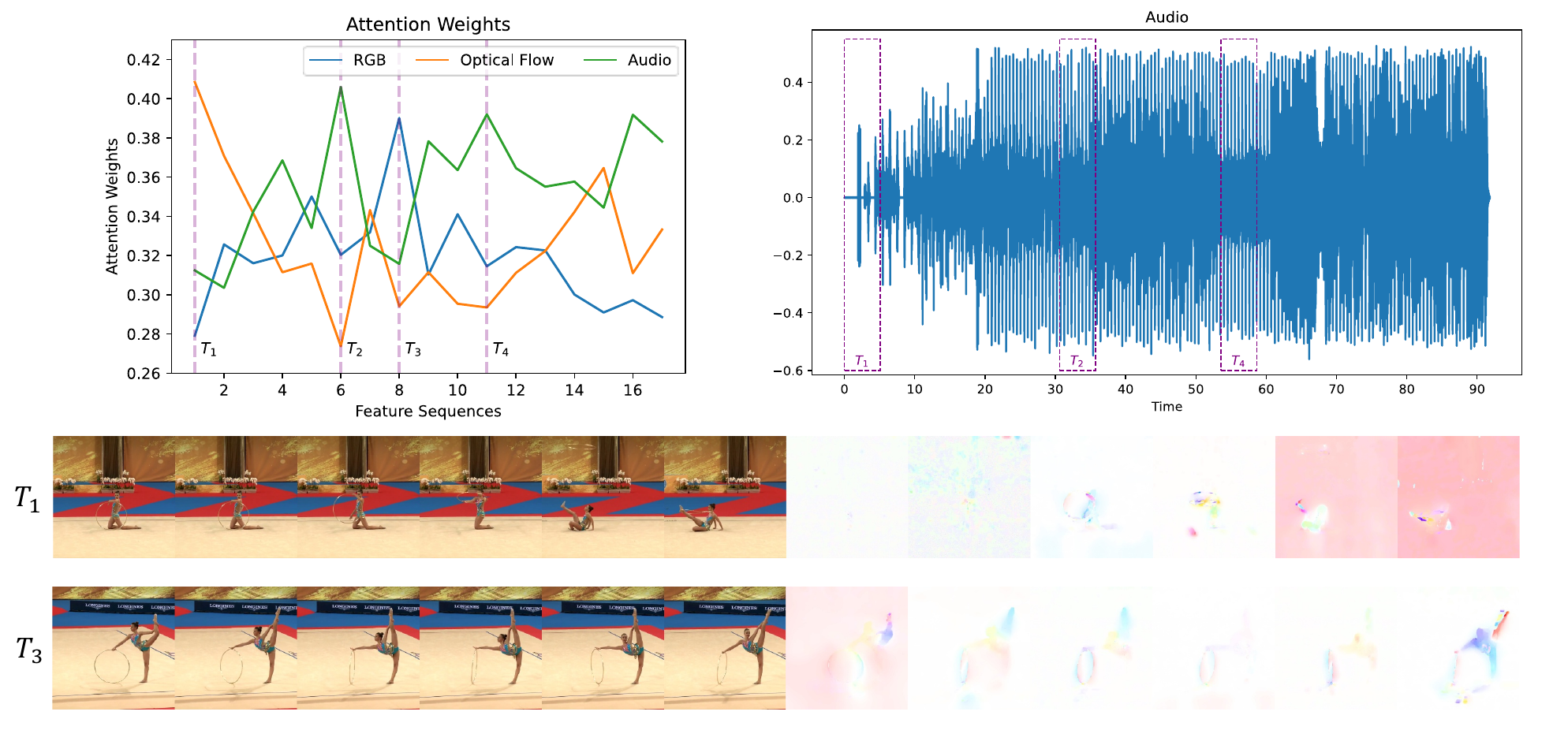}
    \caption{Visualization of attention weights of the Modality-specific Feature Decoder in the second stage on RG (Hoop). The top left figure shows the attention weights of three modalities on feature sequences and four purple vertical dashed lines indicate the time windows corresponding to specific features. The top right figure shows the audio waveform of the video and three rectangles with dashed lines represent the time windows, \ie $T_1$, $T_2$ and $T_4$, in the audio waveform. The next two rows show the RGB frames and optical flows corresponding to the time windows, \ie $T_1$ and $T_3$. Note that each feature in the second stage corresponds to a video segment about 2 seconds and all frames are cropped to keep the gymnast in the center of frames for visualization. This figure is best viewed in color.}
    \label{fig:attention}
    \vspace{-0.4cm}
\end{figure*}

\begin{figure}
    \centering
    \includegraphics[width=\linewidth]{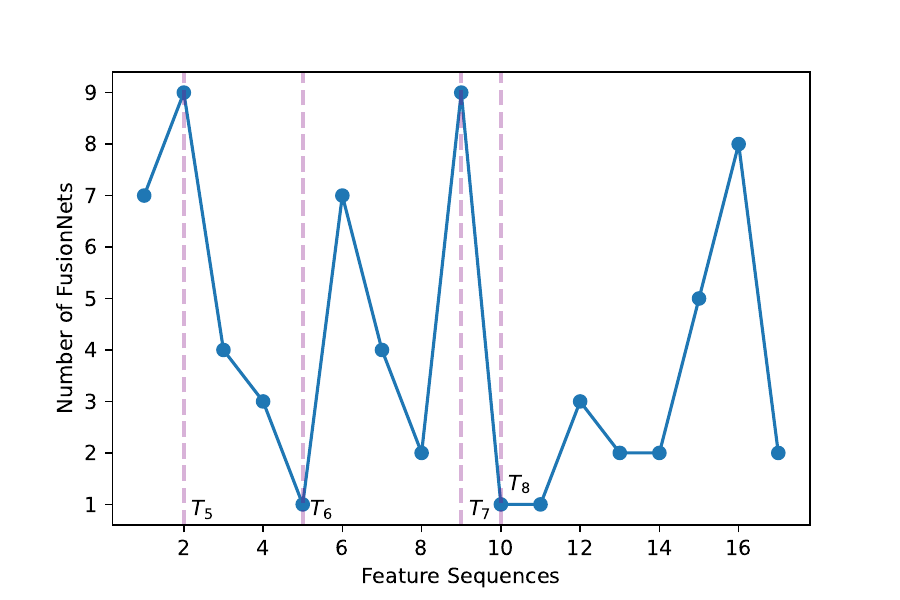}
    \caption{Visualization of decisions generated by PolicyNet in second stage on RG(Hoop). Three purple vertical dashed lines indicate the time windows corresponding to specific features and the corresponding RGB frames, and optical flows are shown in Figure \ref{fig:visual2}. }
    \label{fig:decisions}
    \vspace{-0.3cm}
\end{figure}

\begin{figure*}
    \vspace{-0.5cm}
    \centering
    \includegraphics[width=\linewidth]{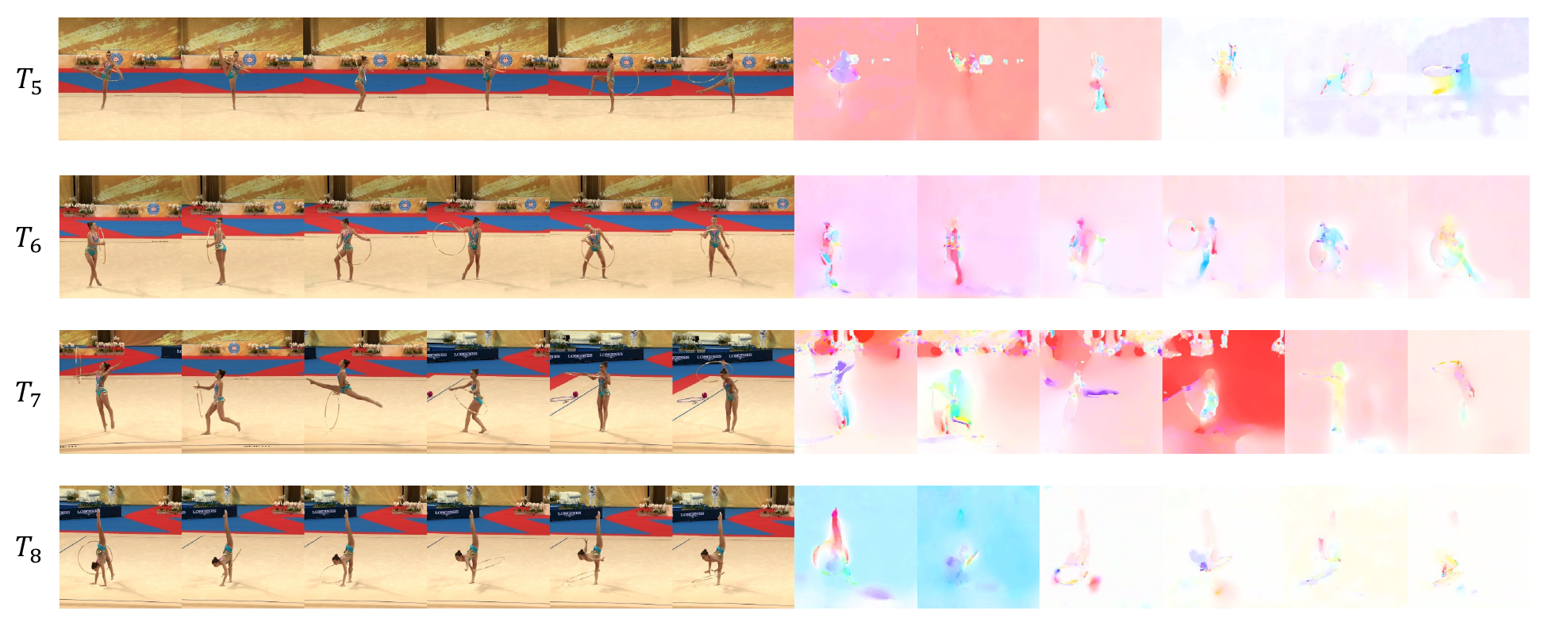}
    \caption{Visualization of RGB frames and optical flows corresponding to the time windows marked in Figure \ref{fig:decisions}, \ie $T_5$, $T_6$, $T_7$ and $T_8$. This figure is best viewed in color. Note that each feature in the second stage corresponds to a video segment about 2 seconds and all frames are cropped to keep the gymnast in the center of frames for visualization.}
    \label{fig:visual2}
    \vspace{-0.3cm}
\end{figure*}

\vspace{0.3cm}

\noindent{\textbf{Implementation Details.}} \an{We use RGB and audio features released by UMT\cite{LiuLWCSQ22}.} For the optical flows, we use I3D \cite{carreira2017quo} pretrained on Kinetics-400 dataset\cite{carreira2017quo} to extract features. The number of FusionNets $K$ is set to 10. We use \an{AdamW \cite{loshchilov2017decoupled}} to optimize our network and train the modality-specific branches and mixed-modality branch for 300 epochs. The batch size is 1 and the learning rate is \an{5e-4}. Others are the same as those used in AQA task.

\vspace{0.3cm}

The experiment results are shown in Figure \ref{tab:hd} and the mean average precision (mAP) is used as the evaluation metric. \an{Our method achieve comparable results to the state-of-the-art method UMT \cite{LiuLWCSQ22} (0.749 vs. 0.747).} Moreover, our method achieves performance improvements of \an{0.066, 0.127 and 0.127} over the modality-specific branches. Remarkably, the modality-specific branch using only audio achieves \an{not bad} performance (\an{0.620}). This result suggests the audio contains rich information for inferring whether a segment is a highlight moment. These results demonstrate the generalizability and effectiveness of our method.}

\subsection{Qualitative Results}

\subsubsection{Visualization of Attention Weights}

To show the importance of different modalities learned by our method, we visualize the attention weights of modality-specific feature decoder on a video feature sequence during the second stage, as shown in Figure \ref{fig:attention}. Additionally, we visualize the audio, video frames and optical flows corresponding to these specific features. As shown in Figure \ref{fig:attention}, since the gymnast always \lab{strikes a pose} and begins to move only when the music starts at the beginning of the video, our model pays more attention to optical flows to assess actions at $T_1$ and we observe that our method usually focuses on optical flows at the beginning of a video on RG dataset. In addition, the audio is assigned to high attention weight at $T_2$ and $T_4$ because $T_2$ includes the chorus of the background music, which has a simple and quick rhythm, and $T_4$ includes a transition from a verse to the chorus. Additionally, when the gymnast rotates the hoop while maintaining body posture at $T_3$, our method focuses on the RGB information to observe the body posture of the gymnast. In summary, we observe that our method uses the background music to assist  in action assessment and focuses on the modality that provides the most useful information for action assessment.

\vspace{0.3cm}

\subsubsection{Visualization of Decisions}
We visualize the decisions generated by the PolicyNet in the second stage on RG (Hoop) in Figure \ref{fig:decisions}, and Figure \ref{fig:visual2} visualizes the RGB frames and optical flows at the four time windows marked in Figure \ref{fig:decisions}. The PolicyNet chooses to use all FusionNets at $T_5$ and $T_7$, and one FusionNet at $T_6$ and $T_8$. As shown in Figure \ref{fig:visual2}, the gymnast turns her body around with leg raised at $T_5$ and performs a split leap at $T_7$, while the gymnast rotates the hoop while twisting or posing at $T_6$ and $T_8$. By comparing the RGB frames and optical flows at $T_5$ and $T_7$ with those at $T_6$ and $T_8$, we find that the PolicyNet tends to use more FusionNets when the gymnast is performing an action with large movements or changing in posture.

\section{Conclusion}
In this work, we propose a novel multimodal method for action quality assessment, called Progressive Adaptive Multimodal Fusion Network (PAMFN), that separately models modality-specific information and cross-modal information. \laa{Our PAMFN consists of three modality-specific branches for independently exploring modality-specific information and a mixed-modality for progressively aggregating the modality-specific information.} Then, to build the bridge between modality-specific branches and the mixed-modality branch, we propose three novel modules. We first propose a Modality-specific Feature Decoder module to selectively transfer modality-specific information to the mixed-modality branch. Then, by taking the potential diversity during different parts of action into consideration, we design an Adaptive Fusion Module to learn adaptive multimodal fusion policies in different parts of an action. Third, we propose a Cross-modal Feature Decoder module to transfer cross-modal features generated by Adaptive Fusion Module to the mixed-modality branch. The state-of-the-art results on two public action quality assessment datasets demonstrate the effectiveness of the proposed model.

\section{Acknowledgment}
This work was supported partially by the NSFC(U21A20471,U1911401), Guangdong NSF Project (No. 2023B1515040025, 2020B1515120085).

\bibliographystyle{IEEEtran}
\bibliography{reference}

\section{Biography Section}

\begin{IEEEbiography}[{\includegraphics[width=1in,height=1.25in,clip,keepaspectratio]{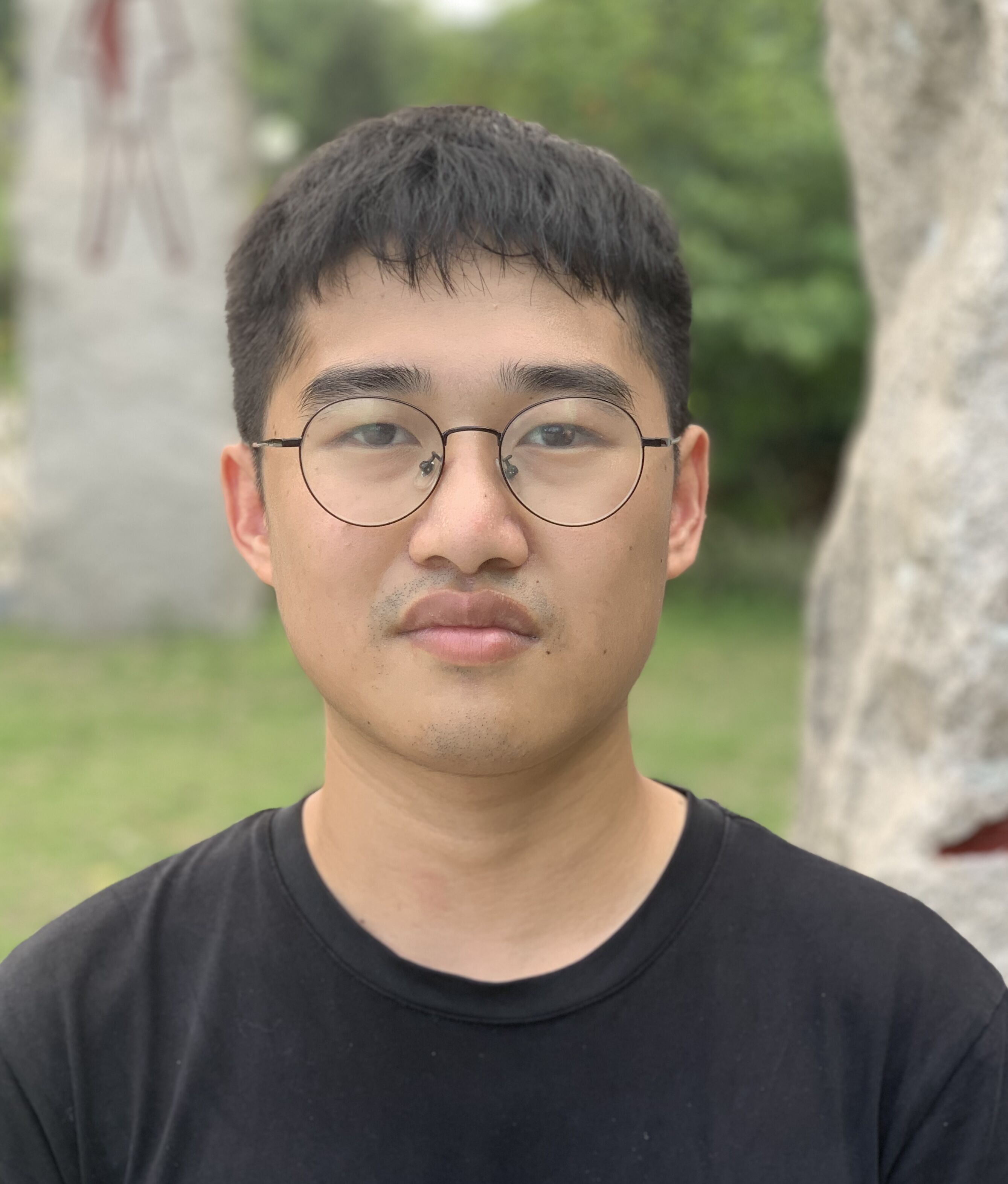}}]{Ling-An Zeng} is received the B.S. degree in Software Engineering from University of Electronic Science and Technology of China in 2019 and the M.S. degree in Computer Technology from Sun Yat-sen University in 2021. He is currently working toward the Ph.D. degree in Computer Science and Technology with the School of Artificial Intelligence in Sun Yat-sen University. His research interests are computer vision and machine learning. He is currently focusing on the topic of action understanding.
\end{IEEEbiography}


\begin{IEEEbiography}[{\includegraphics[width=1in,height=1.25in,clip,keepaspectratio]{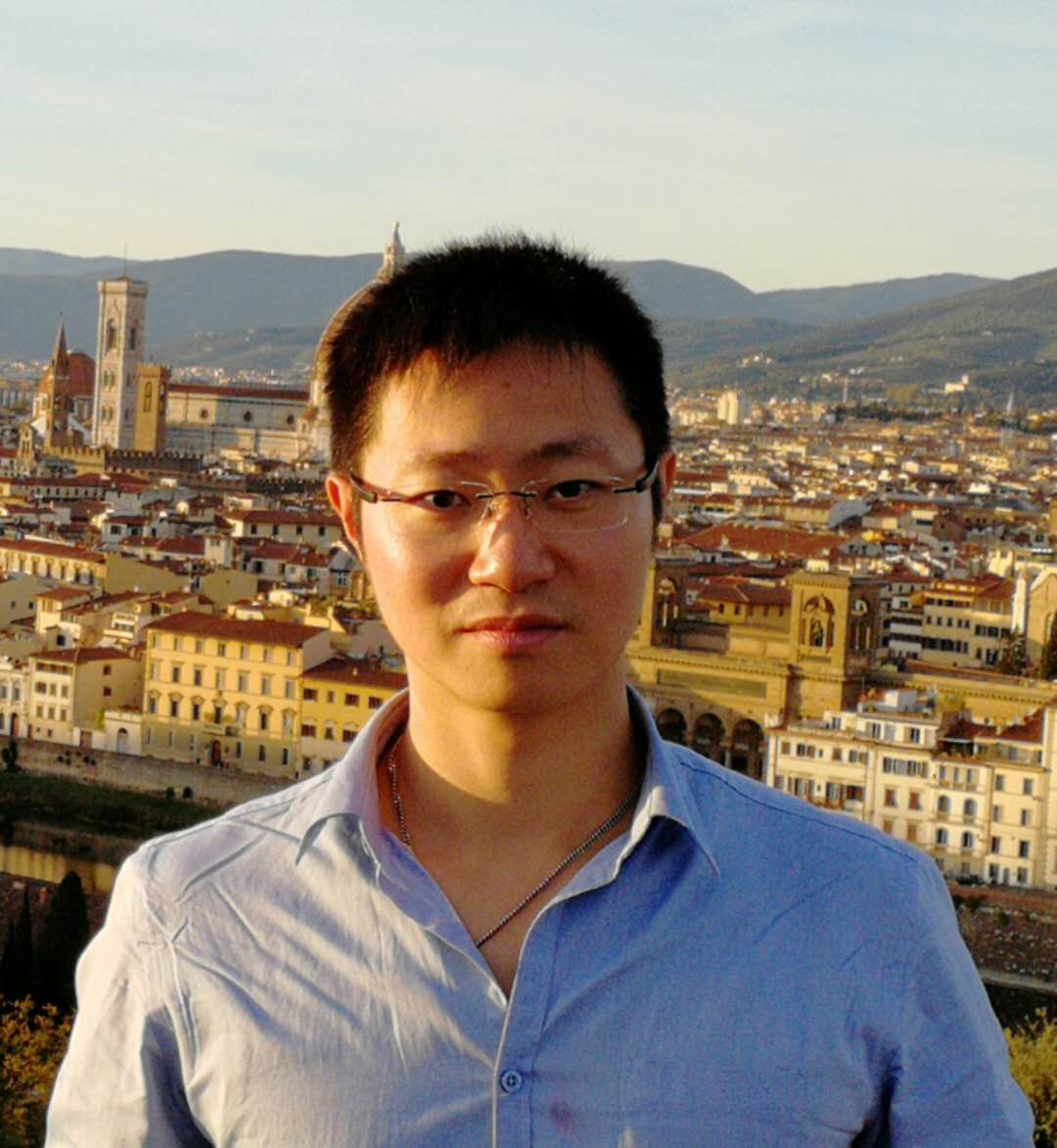}}]{Wei-Shi Zheng} is now a full Professor with Sun Yat-sen University. Dr. Zheng received his Ph.D. degree in Applied Mathematics from Sun Yat-sen University in 2008. His research interests include person/object association and activity understanding in visual surveillance, and the related large-scale machine learning algorithm. Especially, Dr. Zheng has active research on person re-identification in the last five years. He has ever joined Microsoft Research Asia Young Faculty Visiting Programme. He has ever served as area chairs of CVPR, ICCV, BMVC and IJCAI. He is an IEEE MSA TC member. He is an associate editor of the Pattern Recognition Journal. He is a recipient of the Excellent Young Scientists Fund of the National Natural Science Foundation of China, and a recipient of the Royal Society-Newton Advanced Fellowship of the United Kingdom.
\end{IEEEbiography}

\vfill

\end{document}